\documentclass[tightenlines,eqsecnum,floats,aps,amsmath,amssymb,nofootinbib,prd,showpacs,notitlepage]{revtex4-1}
\pdfoutput=1 
\usepackage{amsmath,amssymb,amsfonts}
\usepackage{graphicx}
\usepackage{enumerate} 
\usepackage{colordvi} 
\usepackage{bm}
\usepackage{epsfig}
\usepackage{float}
\usepackage{verbatim}
\usepackage{subfig}
\usepackage{color}

\newcommand{\bfv}{\mbox{\boldmath$v$}}
\newcommand{\bfx}{\mbox{\boldmath$x$}}
\newcommand{\bfk}{\mbox{\boldmath$k$}}

\begin{document}
\title{Modelling Non-Linear Effects of Dark Energy}

\vfill
\author{Benjamin Bose$^{1}$, Marco Baldi$^{2,3,4}$, Alkistis Pourtsidou$^{5,1}$}
\bigskip

\affiliation{$^1$Institute of Cosmology \& Gravitation, University of Portsmouth,
Portsmouth, Hampshire, PO1 3FX, UK}
\affiliation{$^{2}$Dipartimento di Fisica e Astronomia, Alma Mater Studiorum - University of Bologna, Via Piero Gobetti 93/2, 40129 Bologna BO, Italy}
\affiliation{$^{3}$INAF - Osservatorio Astronomico di Bologna, Via Piero Gobetti 93/3, 40129 Bologna BO, Italy}
\affiliation{$^{4}$INFN - Istituto Nazionale di Fisica Nucleare, Sezione di Bologna, Viale Berti Pichat 6/2, 40127 Bologna BO, Italy}
\affiliation{$^{5}$School of Physics and Astronomy, Queen Mary University of London, Mile End Road, London E1 4NS, UK}
\bigskip
\vfill
\date{today}
\begin{abstract}
We investigate the capabilities of perturbation theory in capturing non-linear effects of dark energy. We test constant and evolving $w$ models, as well as models involving momentum exchange between dark energy and dark matter. Specifically, we compare perturbative predictions at 1-loop level against N-body results for four non-standard equations of state as well as varying degrees of momentum exchange between dark energy and dark matter. The interaction is modelled phenomenologically using a time dependent drag term in the Euler equation.  We make comparisons at the level of the matter power spectrum and the redshift space monopole and quadrupole. The multipoles are modelled using the Taruya, Nishimichi and Saito (TNS) redshift space spectrum. We find perturbation theory does very well in capturing non-linear effects coming from dark sector interaction. We isolate and quantify the 1-loop contribution coming from the interaction and from the non-standard equation of state. We find the interaction parameter $\xi$ amplifies scale dependent signatures in the range of scales considered. Non-standard equations of state also give scale dependent signatures within this same regime. In redshift space the match with N-body is improved at smaller scales by the addition of the TNS free parameter $\sigma_v$. To quantify the importance of modelling the interaction, we create mock data sets for varying values of $\xi$ using perturbation theory. This data is given errors typical of Stage IV surveys. We then perform a likelihood analysis using the first two multipoles on these sets and a $\xi=0$ modelling, ignoring the interaction. We find the fiducial growth parameter $f$ is generally recovered even for very large values of $\xi$ both at $z=0.5$ and $z=1$. The $\xi=0$ modelling is most biased in its estimation of $f$ for the phantom $w=-1.1$ case. 
\end{abstract}
\maketitle
\section{Introduction}
The standard model of cosmology (LCDM) -- consisting of dark energy (DE) in the form of a  cosmological constant $\Lambda$ and cold dark matter (CDM) -- has performed extraordinarily well in describing the wealth of precision data the past twenty years has seen. In particular, the best fit model shows a spectacular fit to the latest cosmic microwave background (CMB) measurements by the Planck team \cite{Planck:2015xua}. The model also does very well when compared to low redshift data such as measurements of the baryon acoustic oscillations (BAO) \cite{Anderson:2013zyy} and supernovae data \cite{Riess:2009pu,Lampeitl:2009jq}. On the other hand, from a theoretical standpoint the simple LCDM model has reason to be scrutinized, largely for its $\sim 70\%$ DE component. The old cosmological constant problem \cite{Weinberg:1988cp,Martin:2012bt}, that of fine tuning away the large discrepancy between the observed and predicted vacuum energy, and the new cosmological constant problem, or coincidence problem, are two of the biggest problems in physics. Understanding the current accelerated expansion and probing the nature of dark energy has led  research along two main paths, namely modifications of gravity (MG) and exotic dark energy models (See \cite{Copeland:2006wr,Clifton:2011jh} for reviews). 
\newline
\newline
Recently, LCDM has also come under scrutiny from the observational side too. Discrepancies between the Planck CMB data \cite{Planck:2015xua} and late Universe data using various probes such as clusters \cite{Vikhlinin:2008ym,deHaan:2016qvy}, gravitational lensing \cite{Heymans:2013fya,Abbott:2017wau}, and redshift space distortions (RSD)\footnote{See \cite{Macaulay:2013swa} for a study of the discrepancy between the RSD growth measurements and Planck data.} \cite{Blake:2011rj,Reid:2012sw,Beutler:2013yhm,Gil-Marin:2015sqa,Simpson:2015yfa}, have been uncovered. These discrepancies consistently suggest an overestimation of structure growth when assuming the best-fit Planck LCDM model and evolving it in time, compared to direct low redshift measurements. If confirmed, this consistency seems to support a physical effect, although unknown systematics, such as determination of mass bias in clusters and modelling of non-linear effects in weak lensing, may still be the cause. 
\newline
\newline
If physical, these discrepancies point towards new physics, and one can try to reconcile high and low $z$ measurements by testing models that modify gravity or the energy sector. A common feature of modifications to the gravitational sector is an additional fifth force sourced by extra degrees of freedom, promoting more growth of structure and generally enhancing the discrepancies. Recently however, one avenue that has been promising in relieving these discrepancies is to allow for an interaction within the dark sector \cite{Simpson:2010vh,Lesgourgues:2015wza,Pourtsidou:2016ico,Baldi:2016zom,Buen-Abad:2017gxg,Linton:2017ged} while keeping the theory of gravity that of general relativity (GR). Here we will concentrate on the study of pure momentum exchange interactions between DE and dark matter. Examples of such models are the Dark Scattering model proposed in \cite{Simpson:2010vh}, as well as 
a family of coupled quintessence models constructed in \cite{Pourtsidou:2013nha} using a Lagrangian approach. This type of interaction is qualitatively similar to a drag force on the dark matter particles as they pass through the DE fluid.
\newline
\newline 
In the linear regime these models show promise over the standard fixed $\Lambda$ model in terms of alleviating the structure growth discrepancy \cite{Simpson:2010vh, Pourtsidou:2016ico, Baldi:2016zom}. It is therefore interesting to investigate other detectable signatures of momentum exchange outside the linear regime. Galaxy clustering and redshift space distortions (RSD) \cite{Kaiser:1987qv} offer one such means of distinguishability. RSD is a matter clustering anisotropy that  comes from the non-linear mapping between real position and redshift space position. This mapping must account for the peculiar velocities of the clustering galaxies, which in the presence of momentum exchange, will be damped by the drag force they experience. Signatures of scale dependent, non-linear effects coming from momentum exchange have been measured in the matter power spectrum in N-body simulations \cite{Baldi:2014ica,Baldi:2016zom}. These also employ non-standard equations of state. The magnitude of the deviation of these models from LCDM is of the order of a percent  at $k=0.2 \, h$/Mpc and has a weak redshift dependence. With upcoming spectroscopic surveys such as Euclid \footnote{\url{www.euclid-ec.org}} and DESI \footnote{\url{http://desi.lbl.gov/}}, which aim for sub-percent level accuracy on growth measurements, such a signal could become relevant. Furthermore, the effect momentum exchange has on velocities and the redshift space multipoles, the most relevant observable for spectroscopic survey comparisons, has not been considered. 
\newline
\newline
Simulations are expensive in terms of time and computational power making them ill-suited for statistical comparisons with observational data. For this reason, approximate and phenomenological templates have been widely applied to data, the most common being perturbative techniques \cite{Bernardeau:2001qr,Crocce:2012fa,Taruya:2007xy,Pietroni:2008jx,McDonald:2006hf,Matsubara:2007wj,Bernardeau:2011dp,Blas:2015qsi}. Of these, standard perturbation theory (SPT) (see \cite{Bernardeau:2001qr} for a review) is one of the most popular. These techniques are used to construct the redshift space observables that surveys estimate, the redshift space correlation function and power spectrum being the favoured statistics. 
\newline
\newline
The SPT based Taruya-Nishimichi-Saito (TNS) \cite{Taruya:2010mx} model for the power spectrum is one of the most widely studied and applied theoretical templates. This model does very well in N-body simulation comparisons \cite{Nishimichi:2011jm,Taruya:2013my,Ishikawa:2013aea} and has consequently been employed in deriving constraints on the parameter combination $f \sigma_8$ using survey data \cite{Beutler:2013yhm,Song:2015oza,Beutler:2016arn}, where $f$ is the logarithmic growth rate of structure and $\sigma_8$ gives the normalization of the linear power spectrum. In \cite{Bose:2016qun} a means for consistently constructing the TNS model for general theories of DE and gravity was presented. It is worth noting that the TNS model makes use of a phenomenological damping factor with a single degree of freedom $\sigma_v$. This parameter accounts for small scale velocity dispersions that are beyond truncated SPT's modelling capabilities. The question of whether this degree of freedom is degenerate with non-linear signatures of deviations from the standard model, and more broadly,  the importance of correct theoretical modelling has begun to be addressed in the literature \cite{Taruya:2013my,Barreira:2016mg,Bose:2017myh}. This is of particular importance for upcoming surveys where statistical errors will be minute. By increasing the range of validity of SPT one can place even stronger constraints on modifications and hope to break degeneracies with nuisance parameters. Such extensions are the focus of Effective Field Theory of Large Scale Structures (EFToLSS) \cite{Baumann:2010tm,Carrasco:2012cv,Lewandowski:2015ziq,Perko:2016puo,delaBella:2017qjy}. 
\newline
\newline
In this work we investigate the capabilities of 1-loop SPT to model non-linear effects of modifications to the dark sector in the real and redshift space power spectrum for CDM clustering. Specifically, we focus on the phenomenological Dark Scattering model of \cite{Simpson:2010vh,Baldi:2014ica}. This is done by comparing to a suite of N-body simulations with different models for the equation of state parameter $w$ of DE and varying degrees of interaction within the dark sector. The degeneracy between the interaction strength and $\sigma_v$ is investigated. We organise the paper  as follows: In Sec.~II we review the relevant theory, that is, SPT in momentum exchange models and the TNS template. In Sec.~III we compare the real space predictions for the matter power spectrum for various equations of state of DE and interaction strengths. This is followed by redshift space comparisons, where the focus is on the first two multipoles. This is also done for varying interaction strengths. We quantify the observable signature of the interaction by performing a $\chi^2$ analysis on mock SPT data. Finally, we summarise our results and highlight future work in Sec.~IV.  
\section{Dark Scattering Model: Theoretical Considerations}

The Dark Scattering model we will focus on here was constructed in \cite{Simpson:2010vh}, based on the idea that dark energy and dark matter can interact via elastic scattering. Such an interaction involves only momentum transfer and hence it leaves the density contrast linear perturbation equations unchanged, but introduces a drag term in the velocity perturbations. For the case of cold dark matter, that can be modelled as \cite{Simpson:2010vh,Baldi:2016zom}
\begin{equation}
\dot{\theta}_{\rm c} = -H\theta_{\rm c}
+\frac{\rho_{\rm DE}}{\rho_{\rm c}}(1+w)an_{\rm CDM}\sigma_D(\theta_{\rm DE}-\theta_{\rm c})+k^2\Phi \, ,
\label{eq:theta_scat}
\end{equation}
where $n_{\rm CDM}$ is the proper number density of CDM particles, $\sigma_{\rm D}$ is the DE-CDM scattering cross-section, $\theta_{\rm DE}$ and $\theta_{\rm c}$ are the velocity divergences of the DE and CDM velocity perturbations respectively, $a$ is the scale factor, $w$ is the equation of state for DE and $\Phi$ is the gravitational potential. This interaction can lead to late-time linear growth suppression and is therefore particularly interesting for alleviating the CMB-LSS $\sigma_8$ discrepancy. One can assume that the contribution of the $\delta_{\rm DE}$ interaction term is sub-dominant, a core assumption of the Dark Scattering model. This is convenient for investigating the non-linear effects of such an interaction with N-body simulations, which can only implement dark matter physics. Such Dark Scattering model simulations were performed in \cite{Baldi:2014ica,Baldi:2016zom}, where the assumption that the DE density and velocity fields are approximately homogeneous ($\delta_{\rm DE}=\theta_{\rm DE}=0$) was used. Using this assumption in Eq.~\ref{eq:theta_scat}, the linear Euler equation for CDM can be written as
 \begin{equation}
\dot{\theta}_{\rm c} = -H\theta_{\rm c}
[1+A] + k^2\phi \, ,
\label{eq:theta_cdm}
\end{equation} where 
$A \propto (1+w) \sigma_{\rm D}/m_{\rm CDM}$ at a given time, with $m_{\rm CDM}$ the CDM particle mass. The extra drag (or friction) force introduced by the $A$ term depends on the DE equation of state $w$ and the interaction parameter 
 \begin{equation}
 \xi \equiv \frac{\sigma_{\rm D}}{m_{\rm CDM}}  .
 \label{xidef}
\end{equation}
This is the model we are going to focus on here. We will consider the effects of modifying the background evolution of DE by looking at various DE equation of states as well as the interaction parameter $\xi$, as in \cite{Baldi:2016zom}, and compare with their N-body simulations results. 
\newline
\newline 
It should be noted that a similar kind of interaction, i.e. a pure momentum exchange class of models with no background energy exchange, was constructed using the Lagrangian formalism in \cite{Pourtsidou:2013nha}. These models (dubbed Type 3 models) are a new class of coupled quintessence models and can also be described using the Parameterized Post-Friedmannian framework for interacting dark energy theories developed in \cite{Skordis:2015yra}. An important quantitative difference between the Dark Scattering model and Type 3 models is that the latter predict three extra terms in the velocity equation, proportional to $\theta_{\rm DE}, \theta_{\rm c}$ and the dark energy density contrast $\delta_{\rm DE}$. The latter term is absent in the Dark Scattering models and as shown in \cite{Skordis:2015yra} there is no obvious way to remove its contribution without removing the interaction all-together in these Type 3 models.

\subsection{Background Evolution} 
We begin by assuming a flat Friedman-Robertson-Walker (FRW) background at late times when DM and DE are the only non-negligible energy contributions. Such a background is strongly supported by various experiments, most notably the Planck mission \footnote{http://sci.esa.int/planck/}. The spatial expansion is then described by the Hubble function
\begin{equation}
\left(\frac{H}{H_0}\right)^2 = \left[\Omega_{m,0} a^{-3} + \Omega_{{\rm DE},0}e^{\int^a_13(1+w(a))\tilde{a}}d\tilde{a}\right],
\end{equation}
where $H_0$ is the present day value of the Hubble function, $\Omega_{{\rm DE},0}$ and $\Omega_{m,0}$ are the present day density parameters of DE and DM respectively and $a$ is the scale factor; $w(a)$ is the equation of state parameter of DE. For this work we consider three forms for $w(a)$: constant, the Chevalier-Polarski-Linder form \cite{Chevallier:2000qy,Linder:2002et} and a hyperbolic tangent form \cite{Jaber:2016ucq}
\begin{align}
w_c &\in \{0.9,-1, -1.1\}, \label{wconst}\\ 
w_{\rm CPL} &= w_0 + (1-a)w_a, \label{wcpl} \\  
w_{\rm HYP} &= w_0 +\frac{w_a}{2}[1+\tanh(\frac{1}{a}-1-a_t)], \label{whyp}
\end{align}
where $w_0$ is the parameter at present and $w_a$ sets the time evolution; $a_t$, $w_a$ and $w_0$ are treated as free and we adopt the values presented in Table 1 of \cite{Baldi:2016zom}. None of the chosen models cause more than a $2.5\%$ deviation in the background history when compared to LCDM with the constant models offering most deviation at around $z=1$ (see Fig.~2 of \cite{Baldi:2016zom}). In terms of the drag term $A$, for the observationally relevant redshifts, the constant models also offer the largest magnitude (see Fig.~3 of \cite{Baldi:2016zom}). 

\subsection{Perturbations and Power Spectra}
We now consider perturbations upon the FRW background. In the Newtonian gauge the metric element is given by 
\begin{equation}
ds^2=-(1+2\Phi)dt^2+a^2(1-2\Psi)\delta_{ij}dx^idx^j \, ,
\end{equation} and for the models we study here we have $\Phi = \Psi$.
The evolution equations for matter perturbations are obtained from the conservation of the energy momentum tensor. We assume a DE sound speed of unity, a common prediction for light scalar fields. This damps the DE perturbations within the horizon allowing the safe assumption of the DE to be described as a  homogeneous fluid with $\delta_{\rm DE}=\theta_{\rm DE} =0$ \cite{Baldi:2014ica}. Before shell crossing and assuming no vorticity in the velocity field, a safe assumption at large scales and late times, and in the presence of dark sector momentum transfer,  the evolution equations can then be expressed in Fourier space as \cite{Simpson:2010vh,Baldi:2014ica} 
\begin{eqnarray}
&&a \frac{\partial \delta(\bfk;a)}{\partial a}+\theta(\bfk;a) =-
\int\frac{d^3\bfk_1d^3\bfk_2}{(2\pi)^3}\delta_{\rm D}(\bfk-\bfk_1-\bfk_2)
\alpha(\bfk_1,\bfk_2)\,\theta(\bfk_1;a)\delta(\bfk_2;a) \, ,
\label{eq:Perturb1}\\
&& a \frac{\partial \theta(\bfk;a)}{\partial a}+
\left(2+A+\frac{a H'}{H}\right)\theta(\bfk;a)
-\left(\frac{k}{a\,H}\right)^2\,\Phi(\bfk;a)= \nonumber \\ 
&&\quad \qquad \qquad \qquad \qquad \qquad -\frac{1}{2}\int\frac{d^3\bfk_1d^3\bfk_2}{(2\pi)^3}
\delta_{\rm D}(\bfk-\bfk_1-\bfk_2)
\beta(\bfk_1,\bfk_2)\,\theta(\bfk_1;a)\theta(\bfk_2;a) \, ,
\label{eq:Perturb2}
\end{eqnarray}

where the prime denotes a scale factor derivative, $\delta$ is the density contrast and $\theta$ is the velocity divergence expressed in terms of the peculiar velocity field $\bfv_p({\bfx})$ as $\theta({\bf x})= \frac{\nabla\cdot \bfv_p({\bfx})}{aH(a)}$. The kernels in the Fourier integrals, $\alpha$  and $\beta$, are given by
\begin{eqnarray}
\alpha(\bfk_1,\bfk_2)=1+\frac{\bfk_1\cdot\bfk_2}{|\bfk_1|^2} \, ,
\quad\quad
\beta(\bfk_1,\bfk_2)=
\frac{(\bfk_1\cdot\bfk_2)\left|\bfk_1+\bfk_2\right|^2}{|\bfk_1|^2|\bfk_2|^2} \, .
\label{alphabeta}
\end{eqnarray}
The Newtonian potential $\Phi$ in the Euler equation (Eq.~(\ref{eq:Perturb2})) and its non-linear relation to the matter perturbations  is governed by the Poisson equation \cite{Koyama:2009me}
\begin{equation}
-\left(\frac{k}{a H}\right)^2\Phi=
\frac{3 \Omega_m(a)}{2} \delta(\bfk;a) \, ,
\label{eq:poisson1}
\end{equation}
where $\Omega_m(a) = 8 \pi G \rho_m/3 H^2$. Finally, the term $A$ in the Euler equation is the term coming from momentum exchange. It is given by 
\begin{equation}
A(a) \equiv [1+w(a)]\frac{H_0^2}{H}\frac{3\xi}{8\pi G}\Omega_{DE,0}e^{\int^a_1 \frac{3(1+w(a))}{\tilde{a}} d\tilde{a}} \, ,
\label{interactionterm}
\end{equation}
where $\xi$ is given by Eq.~\ref{xidef} and gives the magnitude of the drag force that will be quoted in units of [bn/GeV]. The term $A$ can act to oppose or enhance the evolution of velocity perturbations depending on whether $w$ is above or below the cosmological constant value $w=-1$. The evolution of this term for the models considered is shown in Fig.~3 of \cite{Baldi:2016zom}. 
\newline
\newline
The assumption of perturbation theory is that the non-linear density and velocity perturbations can be written out as a perturbative expansion of increasing order in the linear perturbations (see \cite{Bernardeau:2001qr} for a review).  Once we assume this, we can solve Eq.~(\ref{eq:Perturb1}) and Eq.~(\ref{eq:Perturb2}) order by order. Specifically we can solve Eq.~(\ref{eq:Perturb1}) and Eq.~(\ref{eq:Perturb2}) perturbatively for \emph{n}-th order kernels $F_n$ and $G_n$ which give the \emph{n}-th order solutions 
\begin{align} 
\delta_n(\boldsymbol{k} ; a) &= \frac{2}{(2\pi)^{3(n-1)}}\int d^3\boldsymbol{k}_1...d^3 \boldsymbol{k}_n \delta_D(\boldsymbol{k}-\boldsymbol{k}_{1...n}) F_n(\boldsymbol{k}_1,...,\boldsymbol{k}_n ; a) \delta_0(\boldsymbol{k}_1)...\delta_0(\boldsymbol{k}_n) \, , \label{nth1} \\ 
\theta_n(\boldsymbol{k}; a) &= \frac{2}{(2\pi)^{3(n-1)}}\int d^3\boldsymbol{k}_1...d^3 \boldsymbol{k}_n \delta_D(\boldsymbol{k}-\boldsymbol{k}_{1...n}) G_n(\boldsymbol{k}_1,...,\boldsymbol{k}_n; a) \delta_0(\boldsymbol{k}_1)...\delta_0(\boldsymbol{k}_n) \, , \label{nth2}
\end{align}
where $\boldsymbol{k}_{1...n} = \boldsymbol{k}_1 + ...+ \boldsymbol{k}_n$. In this work the kernels $F_i$ and $G_i$ are solved for numerically using a modified version of the code described in \cite{Bose:2016qun} which includes the interaction term given in Eq.\ref{interactionterm}. This code employs the algorithm described in \cite{Taruya:2016jdt} which creates the kernels by solving Eq.\ref{eq:Perturb1} and Eq.\ref{eq:Perturb2} iteratively for various values of $\bfk$ in the desired range. For the initial conditions, the solver assumes an Einstein-de Sitter (EdS) cosmology which is valid at early times when the universe was close to matter dominated. This treatment is exact and doesn't assume separability of spatial and temporal components of the kernels. Such separability is known as the EdS approximation as it is an exact treatment in this cosmology. We note that all effects of interacting dark energy on the power spectrum enter through these kernels. At linear order, this is through the linear growth factors $F_1(a)$ (density) and $G_1(a)$ (velocity).  Within the dark scattering models we must solve the following equation 
\begin{equation} 
a F_1''(a)+ \left(3 + A(a) + \frac{aH'(a)}{H(a)}\right) F_1'(a) - \frac{3\Omega_m(a)}{2}F_1(a) = 0, 
\end{equation}
which only differs from a LCDM cosmology through the introduction of the $A$ drag term. There is no scale dependence and so at 1st order, the difference between the perturbations will be through a scale independent enhancement or suppression depending on the sign of $A$. At higher orders the mode coupling terms (right hand side of Eq.\ref{eq:Perturb1} and Eq.\ref{eq:Perturb2}) introduce scale dependencies which will generally differ from LCDM. 
\newline
\newline
For our needs we will expand $\delta$ and $\theta$ up to the third order. Using the perturbations up to third order we can construct the so called 1-loop power spectrum
\begin{equation}
P^{1-{\rm loop}}_{ij}(k;a) = P_{{\rm L},ij}(k;a) + P^{22}_{ij}(k;a) + P^{13}_{ij}(k;a),
\label{loopps}
\end{equation}
where $P_{{\rm L},ij}(k;a)$ is the linear power spectrum defined as 
\begin{equation}
\langle g^1_i(\bfk;a) g^1_j(\bfk';a)\rangle =
(2\pi)^3\delta_{\rm D}(\bfk+\bfk')\,P_{{\rm L},ij}(k;a) \, ,
\end{equation} 
with $\langle ... \rangle$ denoting an ensemble average and where $g^i_\delta = \delta_i$ and $g^i_\theta= \theta_i$. Note that in the models considered in this paper $G_1(a) = -a dF_1(a)/da$ which are generally not unity making the velocity and matter linear spectra not equal, unlike in the Einstein-de Sitter case ($\Omega_m=1$). For brevity, $P_{\rm L}(k;a)$ will refer to the linear matter power spectrum, $P_{{\rm L},\delta \delta}(k;a)$. The higher order terms are given by
\begin{align}
\langle g_i^{2}(\bfk;a) g_{j}^2(\bfk';a)\rangle &=
(2\pi)^3\delta_{\rm D}(\bfk+\bfk')\,P_{ij}^{22}(k;a), \label{eq:psconstraint0} \\
\langle g_i^{1}(\bfk;a) g_{j}^3(\bfk';a)
+g_i^{3}(\bfk;a) g_{j}^1(\bfk';a) \rangle &=
(2\pi)^3\delta_{\rm D}(\bfk+\bfk')\,P_{ij}^{13}(k;a).
\label{eq:psconstraint1}
\end{align}
The inclusion of loop terms has been shown to improve the prediction of theory \cite{Jeong:2006xd}, and generally does better at higher redshift \cite{Carlson:2009it}. Despite this, the loop expansion of the power spectrum is known to have divergent behaviour at small scales making the benefits of the 1-loop terms only enjoyable within a restricted range of scales. We explore the improvement of this first order non-linear extension to the matter power spectrum and redshift space spectrum. The latter we review next.


\subsection{The Redshift Space Power Spectrum}
As we mentioned in the Introduction, the anisotropy of galaxy clustering is directly related to the peculiar velocities of matter. The anisotropy arises from the non-linear mapping between real and redshift space and it is because of the mapping's non-linear nature that makes modelling RSD complex. A model of the effect in the linear regime was first given by Kaiser \cite{Kaiser:1987qv}. It accounts for coherent, linear, infalling motion of galaxies in a cluster
\begin{equation}
P_{{\rm K}}^{\rm (S)}(k,\mu;a) = (1+f\mu^2)^2P_{{\rm L}}(k;a),
\label{linkais}
\end{equation}
where $\mu$ is the cosine of the angle between the line of sight and $\bfk$, $f = d \ln{F_1}/d\ln{a}$ is the logarithmic growth rate of structure, $F_1$ is the linear growth factor as defined in the previous section, and $P_{{\rm L}}(k;a)$ is the linear matter power spectrum. The model does not account for the small scale damping effect of incoherent virialised motion - the fingers-of-God effect. Many authors accounted for this effect via a phenomenological pre factor term, usually taking the form of an Lorentzian or Gaussian \cite{Scoccimarro:2004tg,Percival:2008sh,Cole:1994wf,Peacock:1993xg,Park:1994fa,Ballinger:1996cd,Magira:1999bn}. 
\newline
\newline
For robust tests of gravity, a movement beyond linear models is needed. We consider the TNS model of RSD which is a non-linear semi-perturbative model that has proved its merit through comparisons with simulations, and has consequently been applied to survey data. It is derived partly perturbatively with small scale fingers-of-God effects being treated phenomenologically via a damping factor. The expression for the 2-dimensional redshift space spectrum is given as \cite{Taruya:2010mx}
 \begin{equation}
 P^{\rm (S)}(k,\mu;a) = \mbox{D}_{\mbox{FoG}} (k\mu \sigma_v) \{ P^{1-{\rm loop}}_{\delta \delta} (k;a) - 2  \mu^2 P^{1-{\rm loop}}_{\delta \theta}(k;a) + \mu^4 P^{1-{\rm loop}}_{\theta \theta} (k;a) + A_{\rm TNS}(k,\mu;a) + B_{\rm TNS}(k,\mu;a) \},  
 \label{redshiftps}
 \end{equation}
\noindent where we have absorbed factors of $-f$ into the definition of $\theta$. The $A_{\rm TNS}$ and $B_{\rm TNS}$ terms account for higher-order interactions between the density and velocity fields and are given by 
 \begin{align}
 A_{\rm TNS}(k,\mu;a) &=  -(k \mu) \int d^3 \boldsymbol{k'} \left[  \frac{k_z '}{k'^2} B_\sigma(\boldsymbol{k'},\boldsymbol{k}-\boldsymbol{k'},-\boldsymbol{k};a) +  \frac{k\mu-k_z'}{|\bfk-\bfk'^2|} B_\sigma(\boldsymbol{k}-\boldsymbol{k'}, \boldsymbol{k'},-\boldsymbol{k};a) \right],
 \label{Aterm2} \\
 B_{\rm TNS}(k,\mu;a) &= (k \mu)^2 \int d^3\boldsymbol{k'} Z(\boldsymbol{k'};a) Z(\boldsymbol{k}-\boldsymbol{k'};a),
 \label{Bterm}
 \end{align}
 where
 \begin{equation}
 Z(\boldsymbol{k};a) = \frac{k_z}{k^2}\left[P_{\delta \theta} (k;a) - \frac{k_z^2}{k^2}P_{\theta \theta} (k;a) \right],
 \end{equation}
\noindent and the power spectra here are calculated at linear order to keep the calculation at consistent order with the 1-loop terms. The cross bispectrum $B_\sigma$ is given by
 \begin{equation}
 \delta_D(\boldsymbol{k}_1+ \boldsymbol{k}_2+ \boldsymbol{k}_3)B_\sigma( \boldsymbol{k}_1,\boldsymbol{k}_2,\boldsymbol{k}_3;a) = \langle \theta(\boldsymbol{k}_1;a)\Big\{ \delta(\boldsymbol{k}_2;a) - \frac{k_{2z}^2}{k_2^2} \theta(\boldsymbol{k}_2;a)\Big\}\Big\{ \delta(\boldsymbol{k}_3;a) - \frac{k_{3z}^2}{k_3^2} \theta(\boldsymbol{k}_3;a)\Big\}\rangle.
 \label{lcdmbi}
 \end{equation}
\newline 
We choose an exponential form for  the fingers-of-God damping factor $D_{{\rm FoG}}(k\mu\sigma_v)= \exp{(-k^2 \mu^2 \sigma_v^2)}$, where $\sigma_v$ is treated as a free parameter quantifying the small scale velocity dispersions (expressed in units Mpc/$h$) \cite{Peacock1992}.

\section{Results}
We aim to compare our perturbative approach to the results of full N-body simulations for the Dark Scattering models, that were recently performed by the authors of \cite{Baldi:2016zom} using a suitably modified version of the {\tt GADGET-2} N-body code \cite{Springel:2005mi} that consistently implements the effects of the momentum exchange between dark matter particles and an underlying homogeneous DE field. The simulations evolve an ensemble of $1024^3$ dark matter particles in a periodic cosmological box of $1$ Gpc$/h$ per side, from a starting redshift of $z_{i}=99$ down to the present time. The resulting CDM particle mass is therefore $m_{c} = 8\times 10^{10}$ M$_{\odot }/h$ and the spatial resolution (given by the gravitational softening) is $\epsilon = 24$ kpc$/h$. The simulations cover the set of models summarised in Table 1 of \cite{Baldi:2016zom} for both constant and the CPL-2 and HYP-1 evolving $w$ models, and share the same random phases for the initial conditions realisation of the matter power spectrum, thereby giving rise to the same geometry and topology of the evolved cosmic web at low redshifts and allowing a direct comparison free from cosmic variance. We refer the interested reader to \cite{Baldi:2016zom} for a more extended description of the simulations and of the modified N-body code. Real space Power spectra are computed from simulations snapshots through a Cloud-In-Cell (CIC) mass assignment procedure on a cartesian grid of $1024^{3}$ cells, and for the redshift space spectra physical particle velocities are obtained from the comoving velocities by using the appropriate Hubble function $H(z)$ for each model.
\newline
\newline
First, we compare the 1-loop matter power spectrum with the simulation results. Specifically we are interested in the non-linear modelling of SPT as linear effects of $\xi>0$ and $w(a) \neq -1$ are that of a scale independent enhancement/suppression of power that are degenerate with power spectrum normalization at a given redshift. To see the linear enhancement or suppression of power of the different models when compared to LCDM, we refer the reader to Fig.~4 and Fig.~8 of \cite{Baldi:2016zom}. 

\subsection{Real Space Comparisons}
\begin{figure}[H]
  \captionsetup[subfigure]{labelformat=empty}
  \centering
  \subfloat[]{\includegraphics[width=18cm, height=7.2cm]{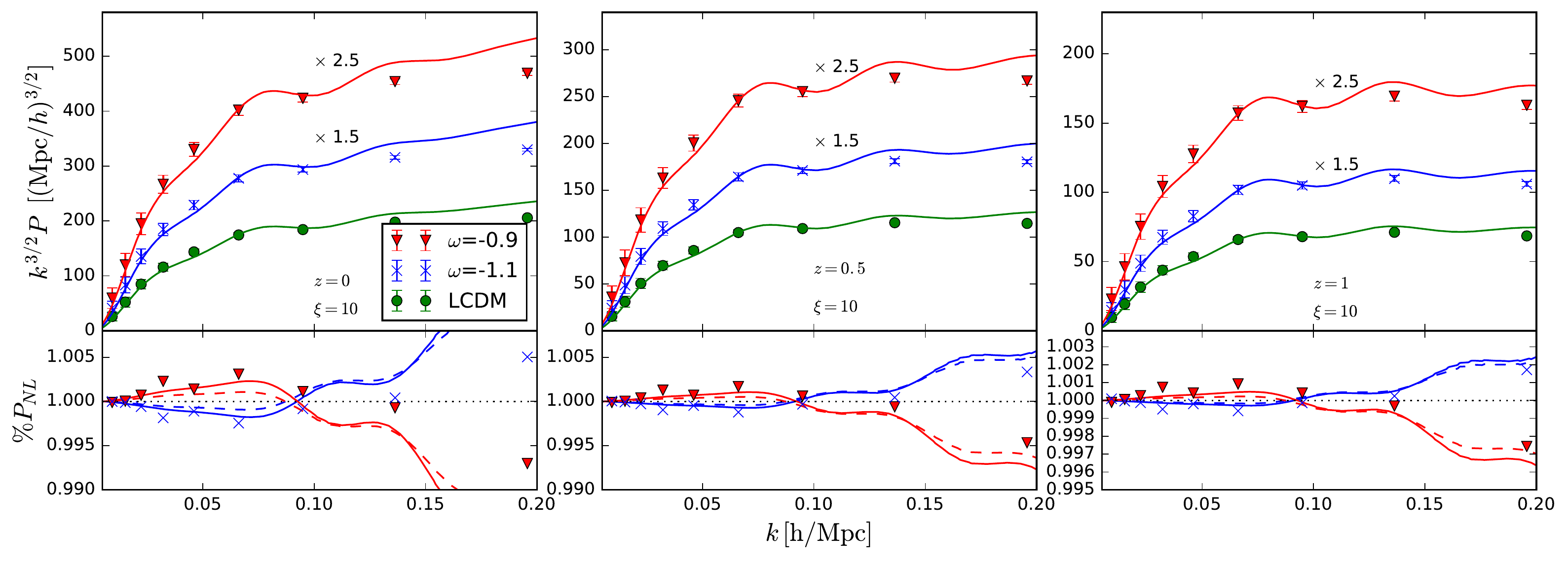}} 
  \caption[CONVERGENCE ]{SPT predictions (solid lines) and N-body measurements (points) of the matter power spectrum in real space at $z=0$ (left), $z=0.5$ (center) and $z=1$ (right) for the fixed $w$ models. The N-body data was fitted with Poisson errors assuming a $1 \, \mbox{Gpc}^3/h^3$ volume and a shot noise term using $\bar{n}= 5 \times 10^{-4} \, h^3/\mbox{Mpc}^{3}$. We have scaled the $w \in \{-1.1,-0.9\}$ spectra for better visualisation. The top panels show the power spectra scaled by $k^{3/2}$ and the bottom panels show the ratio given by Eq.\ref{percentnl}. The dashed lines in the bottom panels show the same quantity for the non-interacting $\xi=0$ case.}
\label{pswfz01}
\end{figure}

\begin{figure}[H]
  \captionsetup[subfigure]{labelformat=empty}
  \centering
  \subfloat[]{\includegraphics[width=18cm, height=7.2cm]{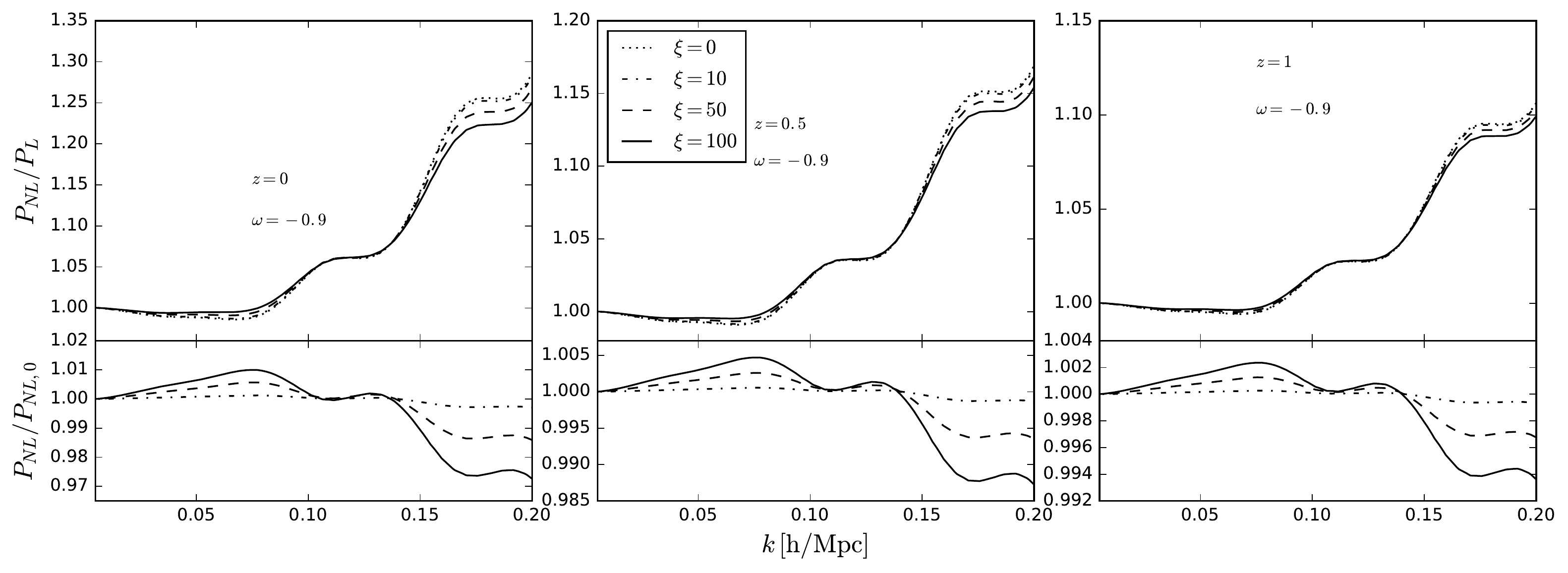}} 
  \caption[CONVERGENCE ]{SPT predictions of the matter power spectrum in real space for various values of $\xi$ at $z=0$ (left), $z=0.5$ (center) and $z=1$ (right) for the $w=-0.9$ case. The top panels show the ratio $P(k)/P_{\rm L}(k)$ and the bottom panels show the ratio of the $\xi=10$, $\xi=50$ and $\xi=100$ curves to the non-interacting $\xi=0$ one.}
\label{pxiw09z01}
\end{figure}

\begin{figure}[H]
  \captionsetup[subfigure]{labelformat=empty}
  \centering
  \subfloat[]{\includegraphics[width=18cm, height=7.2cm]{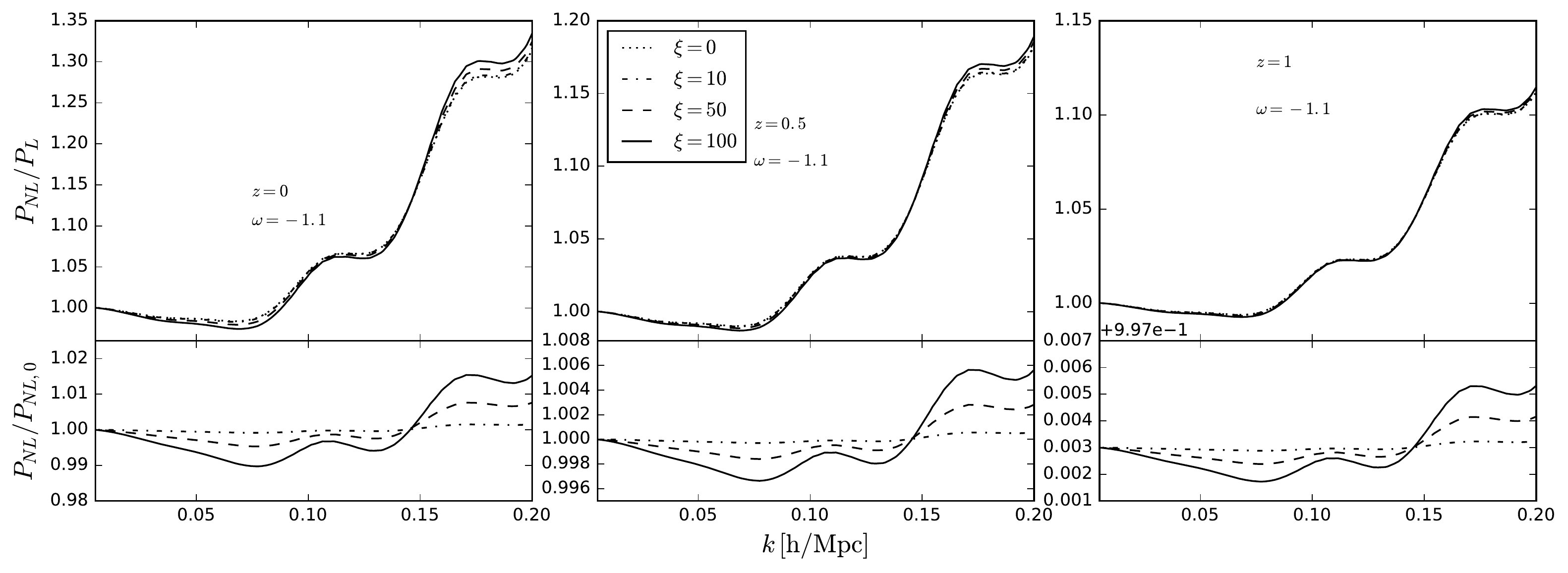}} 
\caption[CONVERGENCE ]{Same as Fig.~\ref{pxiw09z01} for the $w=-1.1$ case.}
\label{pxiw11z01}
\end{figure}

\begin{figure}[H]
  \captionsetup[subfigure]{labelformat=empty}
  \centering
  \subfloat[]{\includegraphics[width=18cm, height=7.2cm]{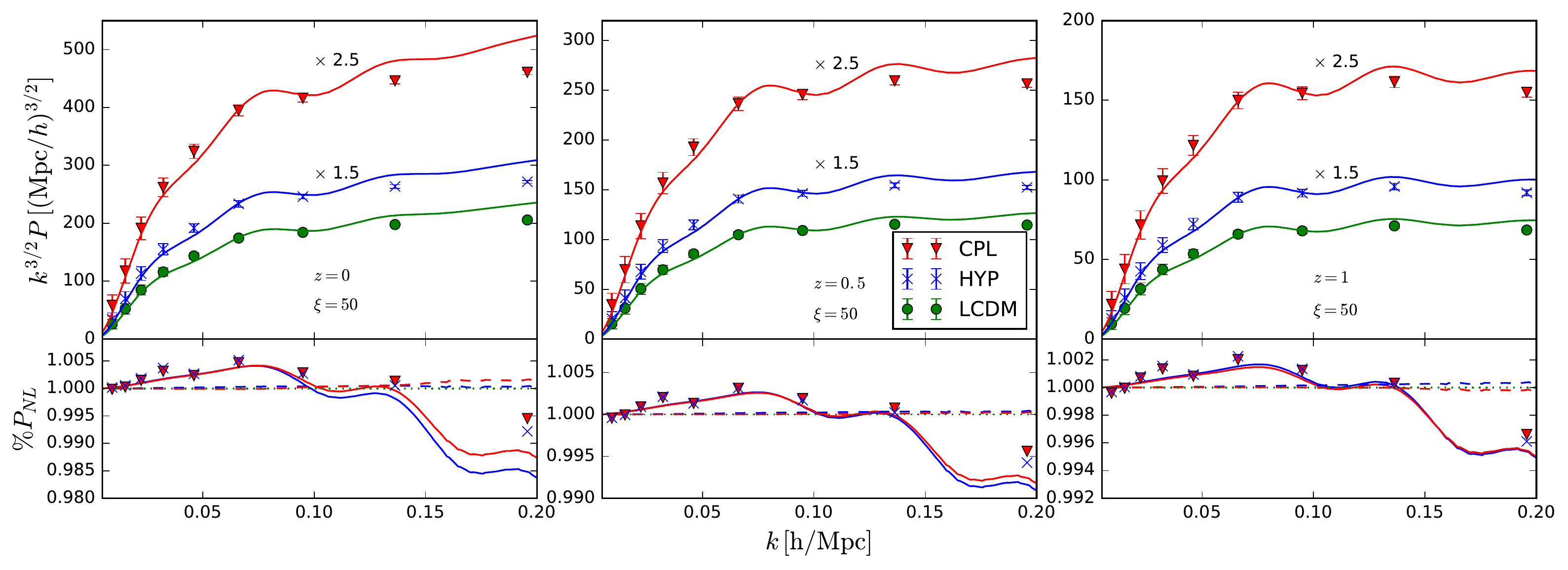}} 
  \caption[CONVERGENCE ]{Same as Fig.~\ref{pswfz01} for the evolving $w$ models. }
\label{pswvz01}
\end{figure}

\begin{figure}[H]
  \captionsetup[subfigure]{labelformat=empty}
  \centering
  \subfloat[]{\includegraphics[width=18cm, height=7.2cm]{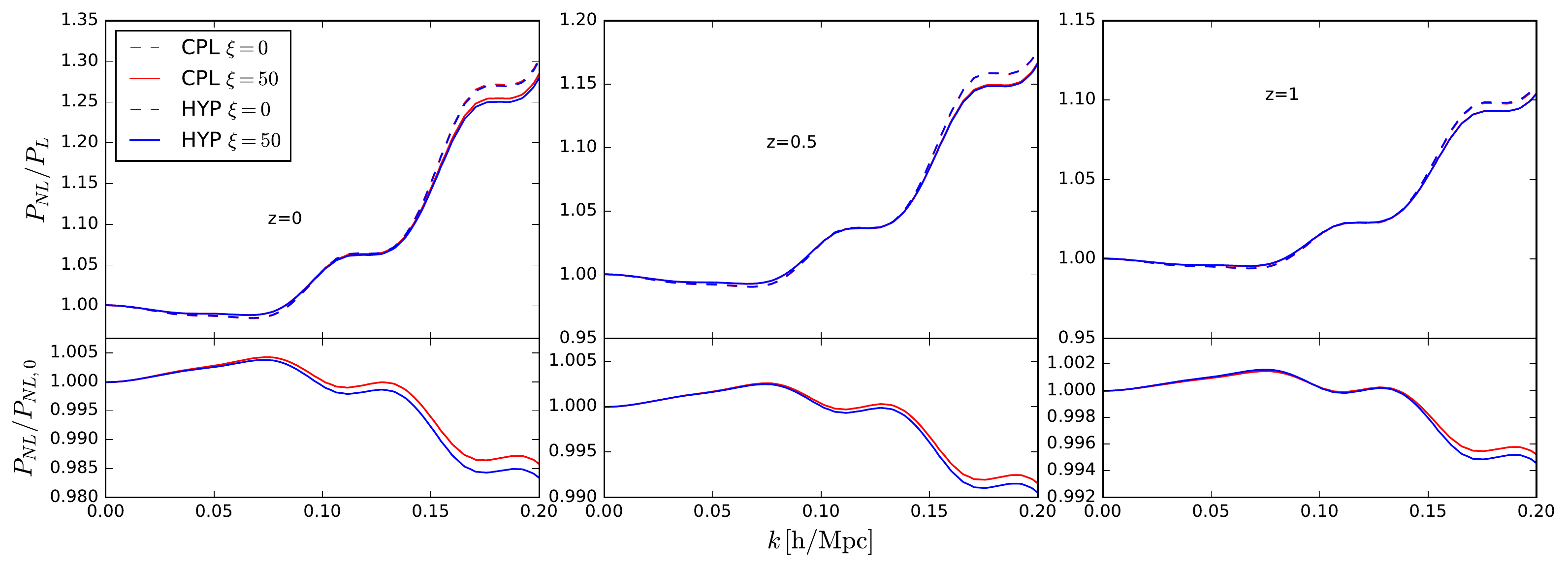}}
\caption[CONVERGENCE ]{SPT predictions of the matter power spectrum in real space for $\xi=0$ and $\xi=50$ at $z=0$ (left), $z=0.5$ (center) and $z=1$ (right) for the evolving $w$ cases. The top panels show the ratio $P(k)/P_{\rm L}(k)$ and the bottom panels show the ratio of the $\xi=50$ curves to the $\xi=0$ ones.}
\label{pxiwvz01}
\end{figure}

The errors shown on the simulation data in this section are computed simply using the number of $k$ modes in each bin and assume a total observed volume of $1 \, \mbox{Gpc}^3/h^3$ (see \cite{Zhao:2013dza} for example). We also include a shot noise term using an average dark matter particle number density of $\bar{n}= 5 \times 10^{-4} \, h^3/\mbox{Mpc}^{3}$.
\newline
\newline
Fig.~\ref{pswfz01} shows the matter power spectra for two interaction models with interaction strength $\xi=10$ and constant $w=-1.1$ and $w=-0.9$. These have been scaled by factors of 1.5 and 2.5 respectively for visualization. The plots also show the LCDM spectrum. We show the results for $z=0$ where non-linearity is maximal and SPT performs worst, and for the higher redshifts $z=0.5$ and $z=1$ that exhibit significantly less non-linearity across the chosen $k$ range. In LCDM, various comparisons with N-body simulations have shown that the 1-loop SPT's $1\%$ deviation range is around $k\sim 0.1 \, h/$Mpc at $z=0$ and $k\sim 0.13 \, h/$Mpc at $z=1$ (see \cite{Carlson:2009it} for example). We find this is consistent with our comparisons in all models.
\newline
\newline
As we work with a single N-body realization for each model, with box size of $1 \, \mbox{Gpc}^3/h^3$, we are subject to large errors at small $k$. All simulations use the same initial seeds so to escape this sample variance, we can use the ratio of each interacting model with LCDM. To quantify the non-linear contribution, we can use the ratio of the `non-linear fraction' of the interaction models to the LCDM one. This quantity is defined as 
\begin{equation}
\% P_{\rm NL} (k) = \frac{P(k)/P_{\rm L}(k)}{P^{\rm LCDM}(k)/P^{\rm LCDM}_{\rm L}(k)}. 
\label{percentnl}
\end{equation} 
\noindent This is shown in the bottom panels of Fig.~\ref{pswfz01}. We see that at large scales, interaction and a value of $w=-0.9$ acts to enhance the power spectrum while a value $w=-1.1$ suppresses power. SPT does very well in modelling this effect at all redshifts with a percent level agreement with simulations all the way up to $k\sim 0.1 \, h$/Mpc. At small scales the effect is the opposite, with $w=-0.9$ acting to suppress power while $w=-1.1$ acting to enhance it. This effect is modelled very well at $z=1$ but is overestimated at low $z$. 
\newline
\newline
In the bottom panels of Fig.~\ref{pswfz01} we also show $\%P_{NL}$ for the $\xi=0$ non-interacting case as dashed lines to disentangle the effect of interaction and the effect of $w \neq -1$. This is shown to be a tiny effect for $\xi=10$ with sub percent effect on the power spectrum at $k\leq 0.1h$/Mpc even at $z=0$. We investigate the effects of enhancing the interaction in Fig.~\ref{pxiw09z01} and  Fig.~\ref{pxiw11z01}.
\newline
\newline
Fig.~\ref{pxiw09z01} and Fig.~\ref{pxiw11z01} show the SPT predictions at $z=0$, $z=0.5$ and $z=1$ for varying levels of interaction quantified by changing $\xi$ from $0$ to $100$ for the constant $w$ cases. The top panels show the ratio of the 1-loop matter power spectrum to the linear matter power spectrum while the bottom panels show the ratio of the 1-loop matter power spectrum with interaction ($\xi \neq 0$) to the non-interacting case ($\xi=0$). This quantifies the non-linear effect coming from enhanced interaction. We emphasise that the linear effect of $w \neq -1$ and $\xi>0$ is that of an overall scaling of $P_L$ and has no $k$-dependence. The non-linear effect is split into two regimes, a large scale enhancement/suppression and small scale suppression/enhancement of power for $w = -0.9/-1.1$ respectively. The level of this effect is $\sim 1\%$ at $k<0.1 \, h$/Mpc at $z=0$ for $\xi=100$. At small scales the effect is more prominent, although at these scales SPT's ability to predict the true power spectrum begins to fail. At these scales, $\xi$ in the $w=-0.9$ case works to reduce the loop contributions while the $w=-1.1$ case works to enhance them. Vice-versa for the larger scales. We also note that the $w=-1.1$ models show less overall effect than their $w=-0.9$ counterparts at small scales. All effects are still sub-percent at $z=1$ for all considered $\xi$ at $k<0.2 \, h$/Mpc. 
\newline
\newline
Next we look at the evolving $w$ models given by Eq.~\ref{wcpl} and Eq.~\ref{whyp}. The cases we restrict our comparisons to are the CPL-2 and HYP-1 models described in Table 1 of \cite{Baldi:2016zom}. For convenience the parameters of these models are $w_0=-1.1$ and $w_a=0.3$ for CPL-2, and $w_0=-1$, $w_a=0.2$ and $a_t=1.5$ for HYP-1. Further $\xi=50$ for both models. We refer to these models simply as CPL and HYP for the remainder of this work.
\newline
\newline
Fig.~\ref{pswvz01} shows the same quantities as Fig.~\ref{pswfz01} but for the CPL and HYP models. The reach of SPT is found to be the same for these models as in LCDM, quantified by the percent deviation from the simulation results. The bottom panels show a very different behaviour to the fixed $w$, $\xi=10$ results with a suppression of power at small scales and a small enhancement at large scales. The order of magnitude of the small scale effects is also similar despite the increased interaction strength with maximum of half a percent signal at $k<0.1 \, h$/Mpc for $z=0$. At all redshifts and all considered scales the effect is sub percent (except $k>0.15 \, h$/Mpc at $z=0$, which is far outside SPT's validity regime). Note that the non-linear effects of the CPL and HYP models are practically indistinguishable.
\newline
\newline
Similarly to the bottom panels of Fig.~\ref{pswfz01}, we show the case of $\xi=0$ as dashed lines. An interesting point is that in these cases we can see that the non-linear effect of the interaction ($\xi>0$) is well captured by SPT, with the dashed $\xi=0$ lines not even within the scatter of the N-body points. Fig.~\ref{pxiwvz01} is similar to 
Fig.~\ref{pxiw09z01} for the CPL and HYP models. The top panels show the 1-loop effects while the bottom panels show the $\xi>0$ effects. Next we look at redshift space. 

\subsection{Redshift Space Comparisons}
\begin{figure}[H]
  \captionsetup[subfigure]{labelformat=empty}
  \centering
  \subfloat[]{\includegraphics[width=18cm, height=8.5cm]{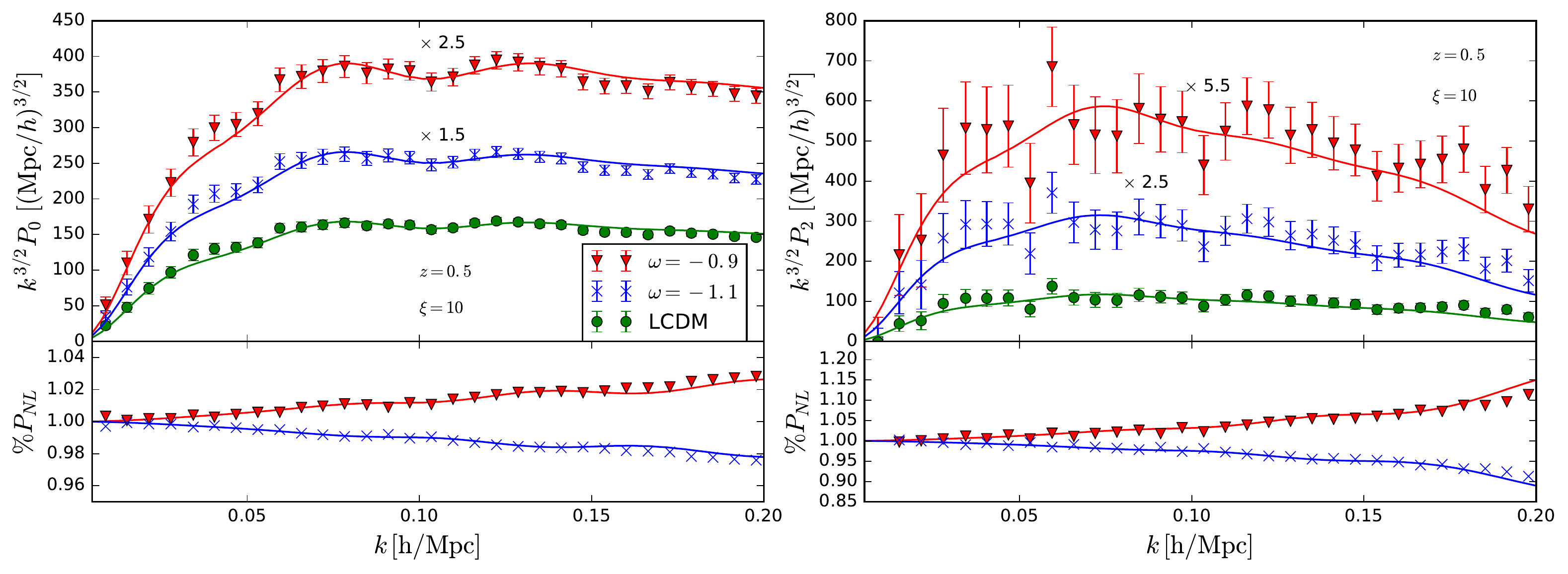}} 
  \caption[CONVERGENCE ]{SPT predictions (solid lines) of the TNS monopole (left) and quadrupole (right) for the fixed $w$ models. The N-body data (points) are fitted with errors coming from the covariance matrix estimate referenced in the text using $V_s=1 \, \mbox{Gpc}^3/h^3$ and $\bar{n}= 5 \times 10^{-4} \, h^3/\mbox{Mpc}^{3}$. We have scaled the $w \in \{-1.1,-0.9\}$ spectra for better visualisation. The top panels show the power spectra scaled by $k^{3/2}$ and the bottom panels show the ratio of  $P_{\ell}(k)/P_{K,\ell}(k)$ to $P^{\rm LCDM}_{\ell}(k)/P^{\rm LCDM}_{K,\ell}(k)$. The best fit $\sigma_v$ ($\chi^2$) are $\sigma_v=4.69 (0.52),4.51(0.56),4.28(0.52)$ for $w=-1.1,-1,-0.9$ respectively. }
 \label{p02wfixz05}
\end{figure}

\begin{figure}[H]
  \captionsetup[subfigure]{labelformat=empty}
  \centering
  \subfloat[]{\includegraphics[width=18cm, height=8.5cm]{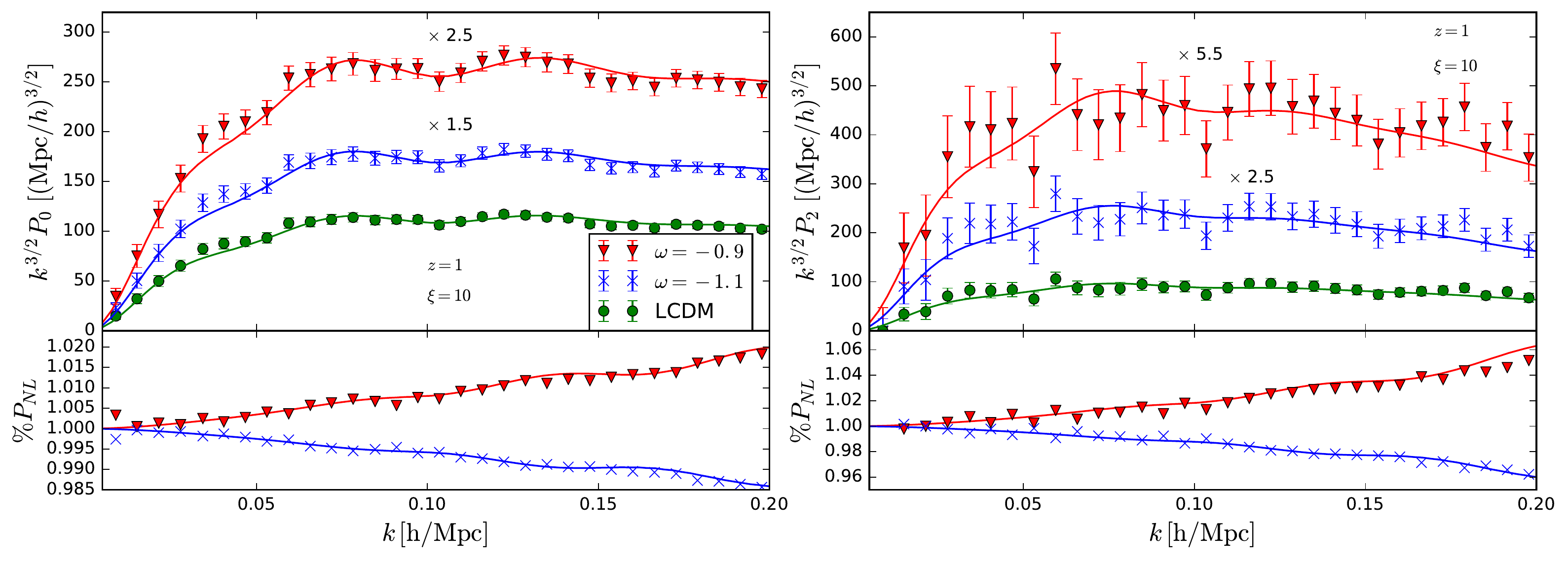}} 
  \caption[CONVERGENCE ]{Same as Fig.~\ref{p02wfixz05} for $z=1$. The best fit $\sigma_v$ ($\chi^2$) are $\sigma_v=3.73 (0.5),3.61(0.49),3.43(0.48)$ for $w=-1.1,-1,-0.9$ respectively. }
\label{p02wfixz1}
\end{figure}

\begin{figure}[H]
  \captionsetup[subfigure]{labelformat=empty}
  \centering
  \subfloat[]{\includegraphics[width=18cm, height=8.5cm]{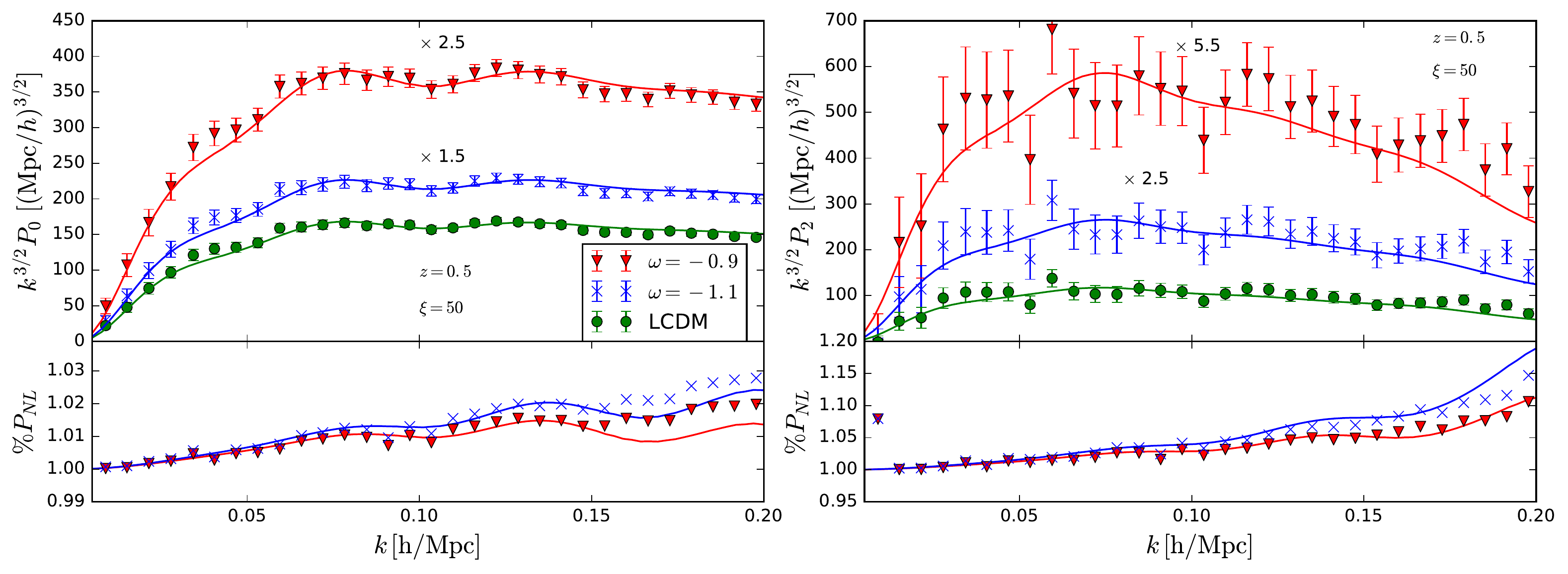}}  
  \caption[CONVERGENCE ]{Same as Fig.~\ref{p02wfixz05} for the CPL (red), HYP (blue) and LCDM (green) models at $z=0.5$ and $\xi=50$. The best fit $\sigma_v$ ($\chi^2$) are $\sigma_v=4.27 (0.52),4.36(0.52)$ for the HYP and CPL models respectively. See Fig.~\ref{p02wfixz05} for the LCDM value.}
 \label{p02wvarz05}
\end{figure}

\begin{figure}[H]
  \captionsetup[subfigure]{labelformat=empty}
  \centering
  \subfloat[]{\includegraphics[width=18cm, height=8.5cm]{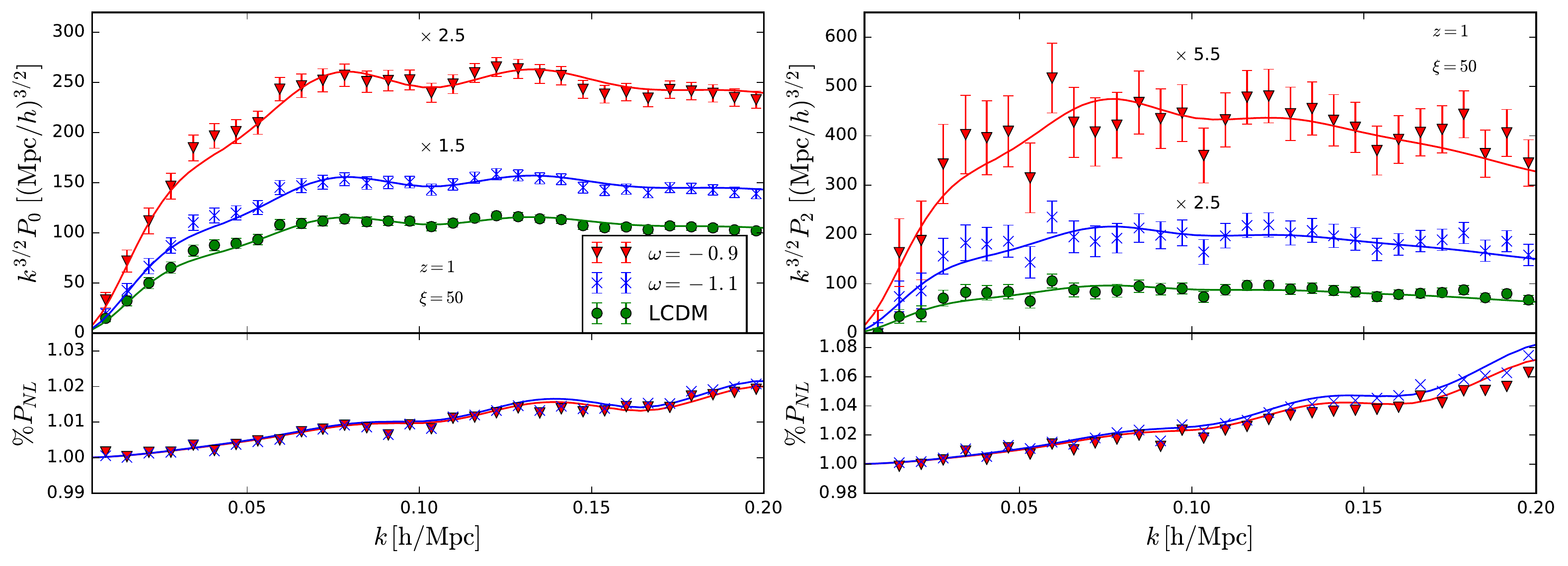}} 
  \caption[CONVERGENCE ]{Same as Fig.~\ref{p02wfixz05} for the CPL, HYP and LCDM models at $z=1$ and $\xi=50$. The best fit $\sigma_v$ ($\chi^2$) are $\sigma_v=3.4 (0.48),3.42(0.47)$ for the HYP and CPL models respectively. See Fig.~\ref{p02wfixz1} for the LCDM value.}
 \label{p02wvarz1}
\end{figure}

\begin{figure}[H]
  \captionsetup[subfigure]{labelformat=empty}
  \centering
  \subfloat[]{\includegraphics[width=18cm, height=8.5cm]{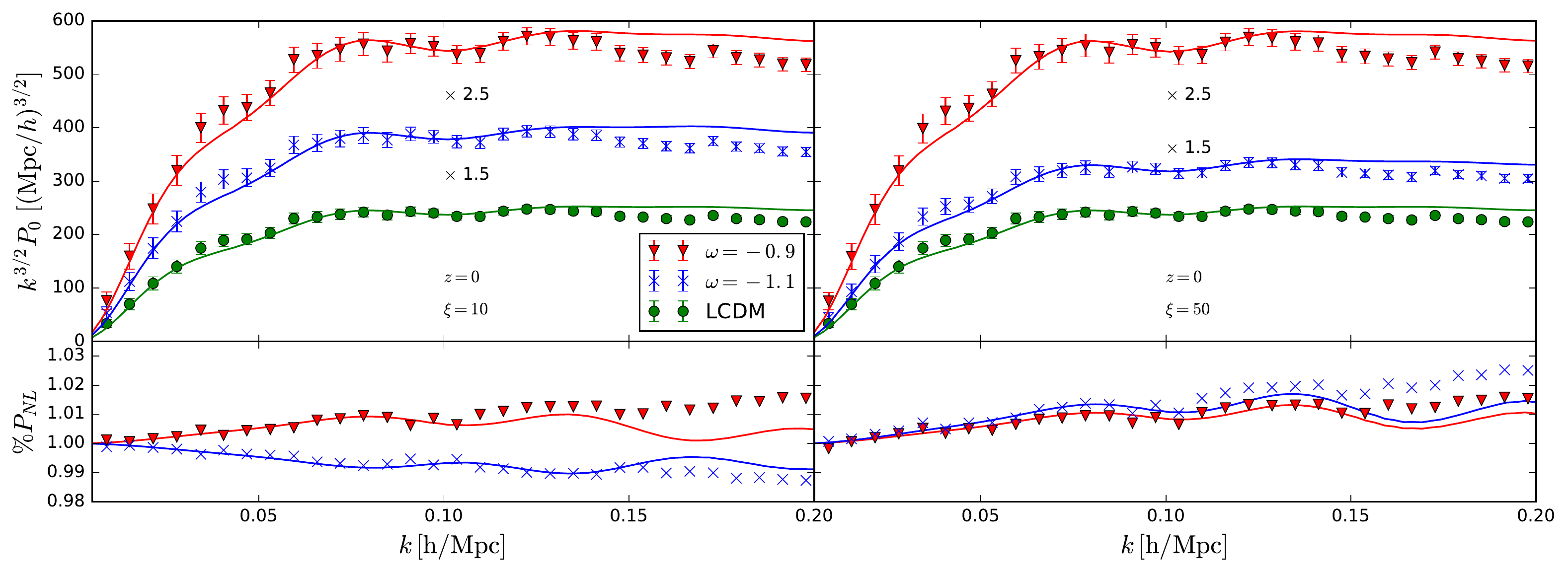}}
  \caption[CONVERGENCE ]{The monopoles for the fixed (left) and evolving (right) $w$ models at $z=0$. The colour scheme is the same as used throughout this section. }
 \label{p0z0}
\end{figure}

In this section we look at the redshift space monopole and quadrupole. Simulations with larger volumes and high mass resolution are required to model higher order multipoles and so we omit them here. These are computed from the simulation snapshots by virtually placing the simulation box at the appropriate comoving distance from the observer and changing the observed redshift along each of the three cartesian axes of the box by adding to the Hubble expansion of the universe the component of the peculiar velocities of particles along that particular axis. This procedure, known as the {\em distant observer approximation}, provides three correlated realisations of the observed redshift space spectrum so that a simulated measurement can be obtained by averaging over them. This puts the effective volume related to the measurement between $1 \mbox{Gpc}^3/h^3$ and $3  \mbox{Gpc}^3/h^3$. We opt to choose the conservative value of $1 \mbox{Gpc}^3/h^3$ in error calculations as our core results involve ratios of the GR and dark scattering model measurements where cosmic variance is significantly reduced. The multipoles are modelled as 
\begin{equation}
P_\ell^{(S)}(k)=\frac{2\ell+1}{2}\int^1_{-1}d\mu P_{TNS}^{(S)}(k,\mu)\mathcal{P}_\ell(\mu),
\end{equation}
where $\mathcal{P}_\ell(\mu)$ denote the Legendre polynomials and $P_{TNS}^{(S)}(k)$ is given by Eq.~(\ref{redshiftps}). Further, we only consider $z=0.5$ and $z=1$ as $z=0$ is observationally irrelevant for spectroscopic surveys. Fig.~\ref{p02wfixz05} and Fig.~\ref{p02wfixz1} show the N-body measurements against the constant $w$ SPT predictions for $z=0.5$ and $z=1$ respectively. The top panels show excellent agreement at $k<0.2 \, h$/Mpc, with the $w \in \{ -1.1,-0.9\}$ models scaled for visualization. The fit is aided by the free parameter $\sigma_v$ which we fit to the data by minimizing the $\chi^2$ given by 
\begin{equation}
\chi^2 = \frac{1}{2N_k - N_p} \sum_n \sum_{\ell,\ell'=0,2} \left(P^{(S)}_{\ell,{\rm data}}(k_n)-P^{(S)}_{\ell,{\rm model}}(k_n)\right) \mbox{Cov}^{-1}_{\ell,\ell'}(k_n)\left(P^{(S)}_{\ell',{\rm data}}(k_n)-P^{(S)}_{\ell',{\rm model}}(k_n)\right),
\label{covarianceeqn}
\end{equation}
with $N_k$ being the number of $k$-bins we consider, $N_p=1$ is the number of free parameters and $k_n$ being the $n$th bin. $\mbox{Cov}_{\ell,\ell'}$ is the covariance matrix between the different multipoles. Expressions for the covariance components can be found in Appendix C of \cite{Taruya:2010mx}. Again, we use a volume of $V_s=1 \, \mbox{Gpc}^3/h^3$ and include the effect of shot-noise assuming a dark matter particle density of $\bar{n}= 5 \times 10^{-4} \, h^3/\mbox{Mpc}^{3}$. Note that we use each model's linear theory to estimate the covariance matrix components \footnote{The linear predictions vary with $\xi$ and $w$.}. This approximation has been checked to work well within $k\leq 0.3 \, h$/Mpc for the LCDM simulations used in \cite{Taruya:2010mx}. Because at large scales the scale dependent modifications to LCDM we consider are so small, we feel this is valid for our purposes of simply fitting $\sigma_v$. Further, we point out that the effective volume is actually slightly larger than $1 \mbox{Gpc}^3/h^3$ as we average over 3 correlated measurements. This will increase the $\chi^2$ slightly. We use 24 $k$ bins for $z=1$, fitting up to $k_{\rm max} = 0.15 \, h$/Mpc, and 20 bins for $z=0.5$ fitting up to $k_{\rm max} = 0.13 \, h$/Mpc. 
\newline  
\newline 
\newline 
As we have already mentioned, the quadrupole is very noisy for reasons of limited volume and resolution. To get around the noise and isolate non-linear effects, we again plot the quantity of Eq.~(\ref{percentnl})\footnote{$P_L$ is now given by the multipoles of Eq.~\ref{linkais}.} in the bottom panels. Again the TNS+SPT modelling does very well at capturing non-linear effects. The $w=-0.9/-1.1$ model shows a steadily increasing/decreasing enhancement/suppression of both multipoles with respect to LCDM. We make the note that there are features in these effects and they are not strictly monotonic. This is shown in Fig.~\ref{p0z0}, which is the $z=0$ case. At higher redshift these features are less developed and may be enhanced by $\xi$. We discuss this in the next section.  
\newline 
\newline
Fig.~\ref{p02wvarz05} and Fig.~\ref{p02wvarz1} show the same results for the CPL and HYP models. Again we find very good agreement with N-body. The bottom panels also show the non-linear features being modelled well by SPT for $k<0.15 \, h$/Mpc for $z=0.5$ and $k<0.2 \, h$/Mpc for $z=1$. We note oscillatory features in the bottom panels. These non-linear effects will be key in breaking the degeneracy with $\sigma_v$ and other nuisance parameters. On this note, at this stage it is much harder to disentangle the effect of $\xi>0$ with $w \neq -1$ as we have introduced the additional degree of freedom $\sigma_v$. We attempt to address this point in the next Section.

\newpage
\subsection{Significance of Interaction in Future Surveys}
\begin{figure}[H]
  \captionsetup[subfigure]{labelformat=empty}
  \centering
  \subfloat[]{\includegraphics[width=18cm, height=7.2cm]{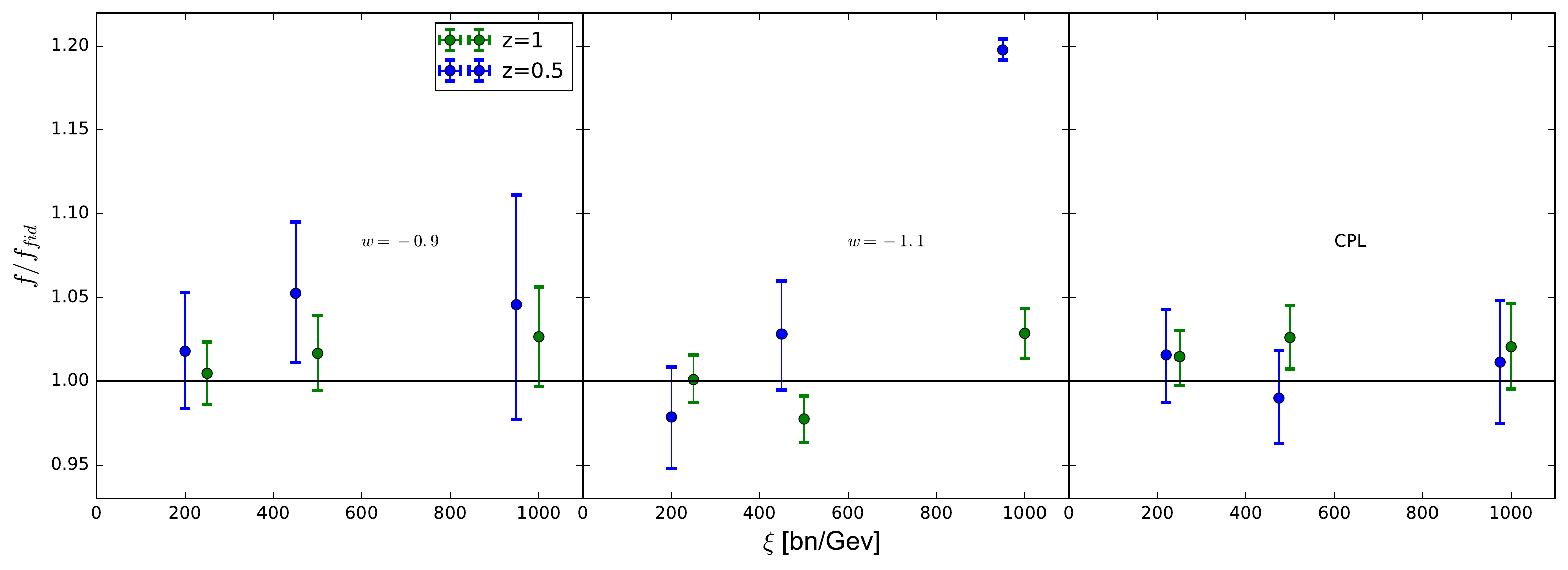}} 
\caption[CONVERGENCE ]{The marginalized best fit growth as a function of $\xi$ used in constructing the SPT mock data sets. This is shown for the $w=-0.9$ (left), $w = -1.1$ (center) and CPL (right) cases with the attached $2\sigma$ error bars. The mock data assumes ideal survey errors with $V_s=20 \, \mbox{Gpc}^3/h^3$ for $z=1$ and $V_s=10 \, \mbox{Gpc}^3/h^3$ for $z=0.5$, with a shot noise term of $\bar{n}=5\times 10^{-3} \, h^3/\mbox{Mpc}^3$.  The $z=0.5$ data points have been shifted slightly for better visualization.}
\label{frecovery1}
\end{figure}

\begin{figure}[H]
  \captionsetup[subfigure]{labelformat=empty}
  \centering
  \subfloat[]{\includegraphics[width=8.7cm, height=8.7cm]{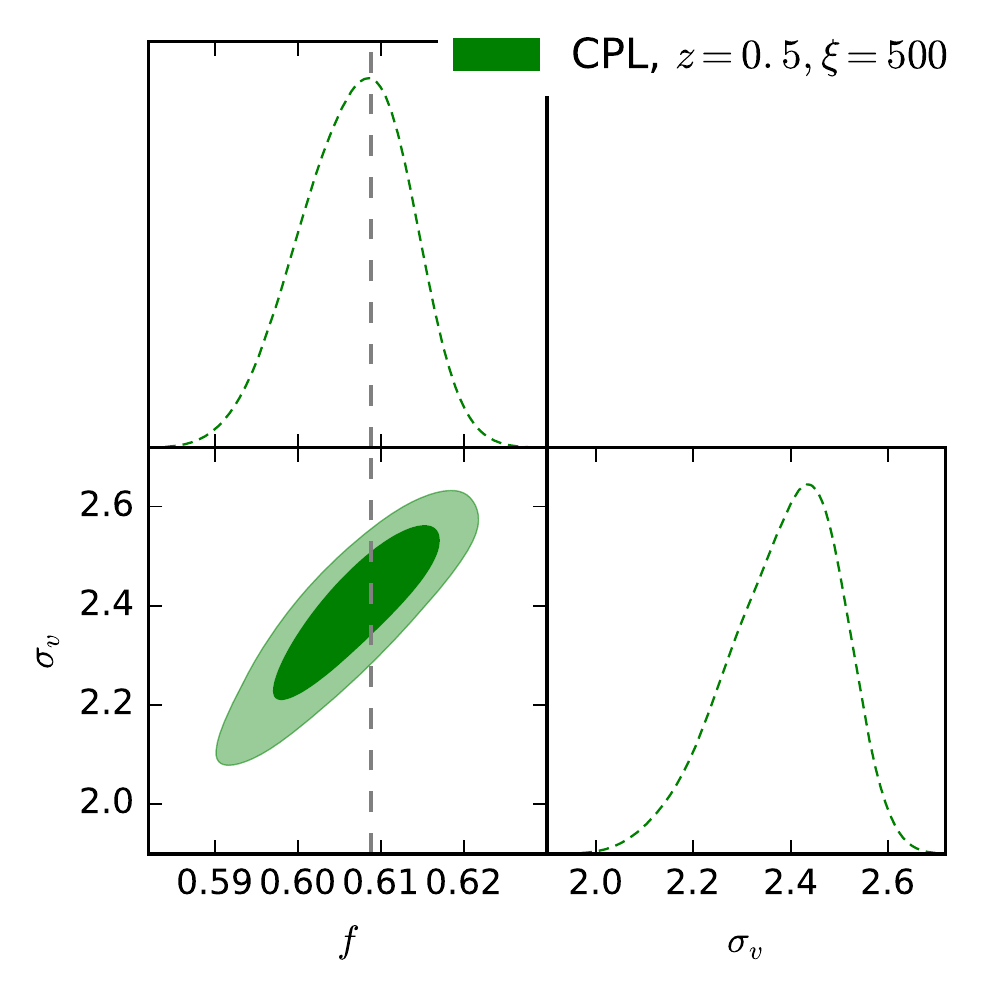}} \quad
   \subfloat[]{\includegraphics[width=8.7cm, height=8.7cm]{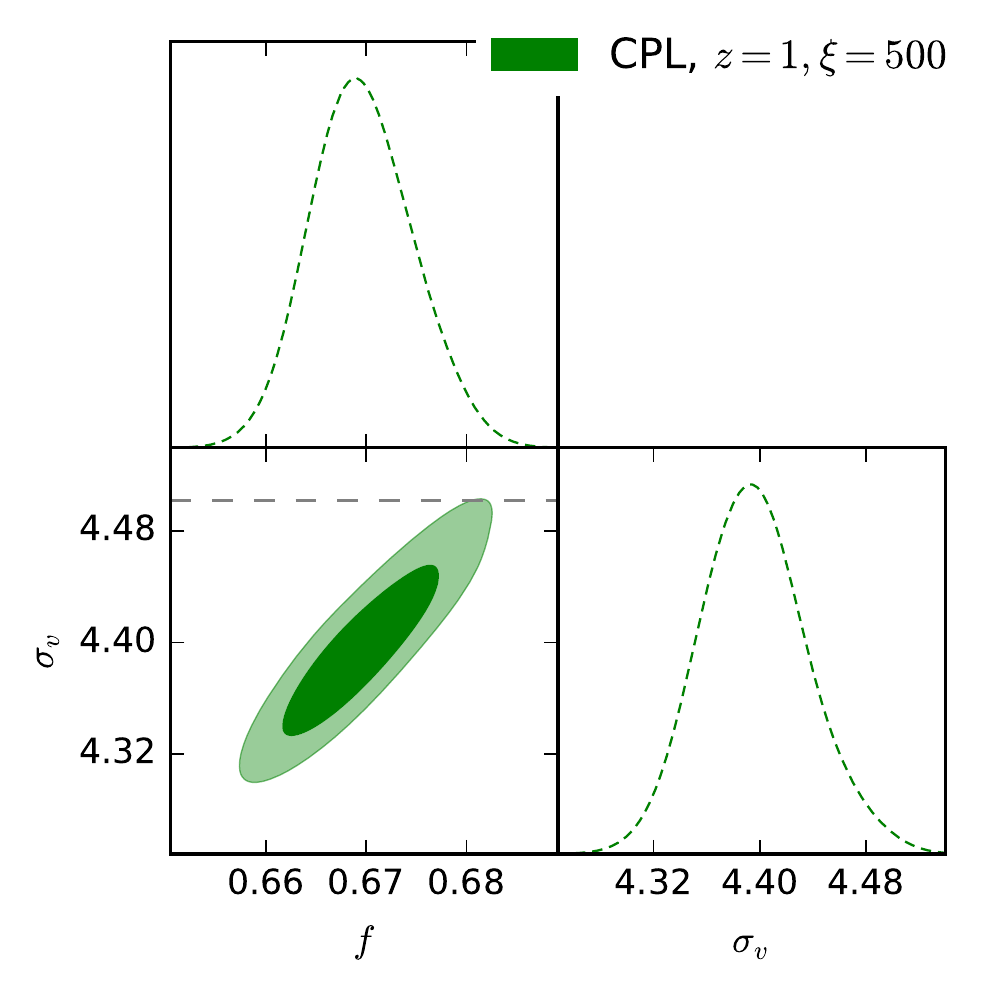}} 
\caption[CONVERGENCE ]{The 2D Likelihood contours ($68\%$ CL and $95\%$ CL) on $\sigma_v$ and $f$ for the $\xi=500$, CPL case for $z=0.5$ (left) and $z=1$(right). The fiducial parameters are $\sigma_v=3,4.5 \mbox{Mpc}/h$ and $f=0.608,0.650$ for $z=0.5$ and $z=1$ respectively.}
\label{frecovery2}
\end{figure}

\begin{figure}[H]
  \captionsetup[subfigure]{labelformat=empty}
  \centering
  \subfloat[]{\includegraphics[width=18cm, height=8.7cm]{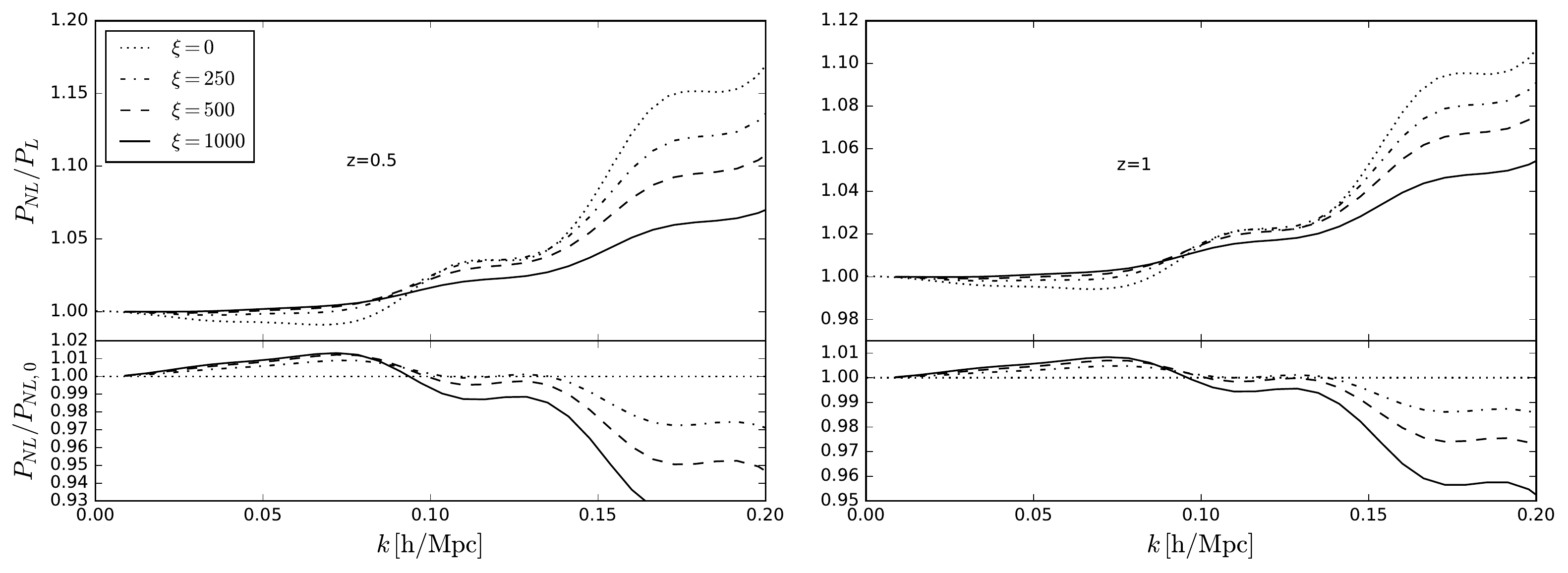}} 
\caption[CONVERGENCE ]{SPT predictions of the matter power spectrum in real space for various values of large $\xi$ at $z=0.5$ (left) and $z=1$ (right) for the $w=-0.9$ case. The top panels show the ratio $P(k)/P_{\rm L}(k)$ and the bottom panels show the ratio of the $\xi \neq 0$ curves to the $\xi=0$ one.}
\label{pnl_w09}
\end{figure}

\begin{figure}[H]
  \captionsetup[subfigure]{labelformat=empty}
  \centering
  \subfloat[]{\includegraphics[width=18cm, height=8.7cm]{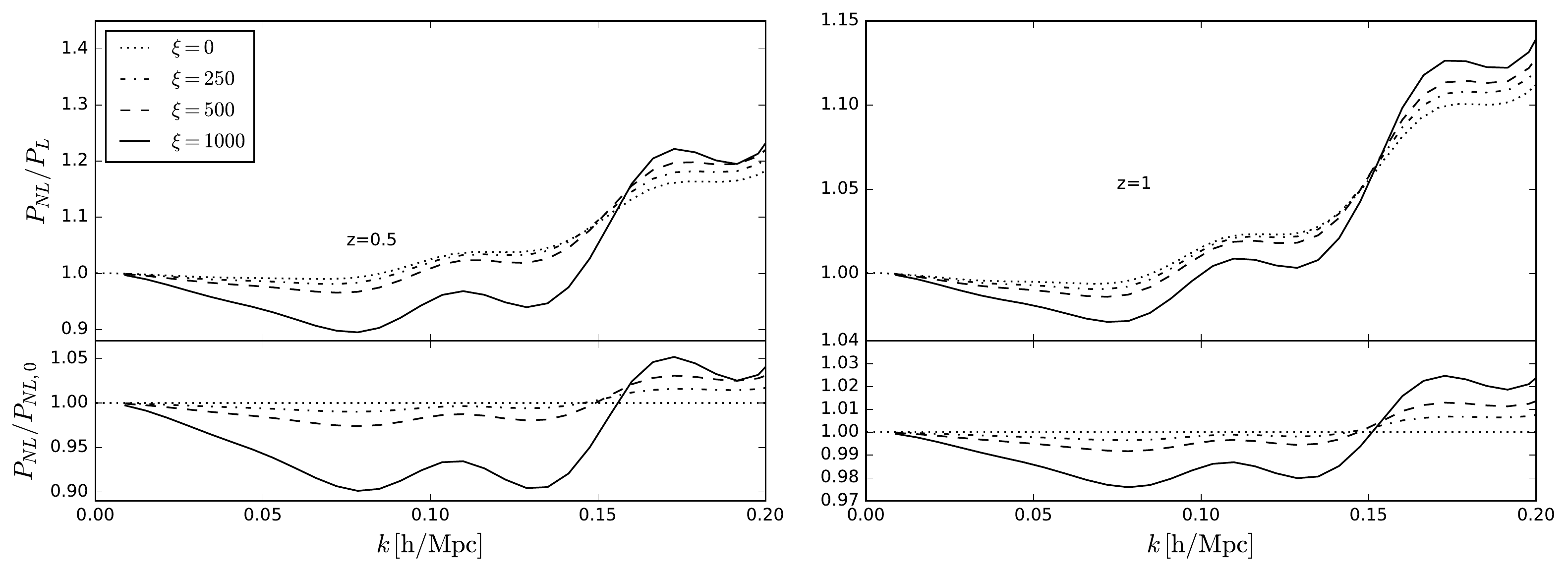}} 
\caption[CONVERGENCE ]{Same as Fig.~\ref{pnl_w09} for $w = -1.1$.}
\label{pnl_w11}
\end{figure}

\begin{figure}[H]
  \captionsetup[subfigure]{labelformat=empty}
  \centering
  \subfloat[]{\includegraphics[width=18cm, height=8.7cm]{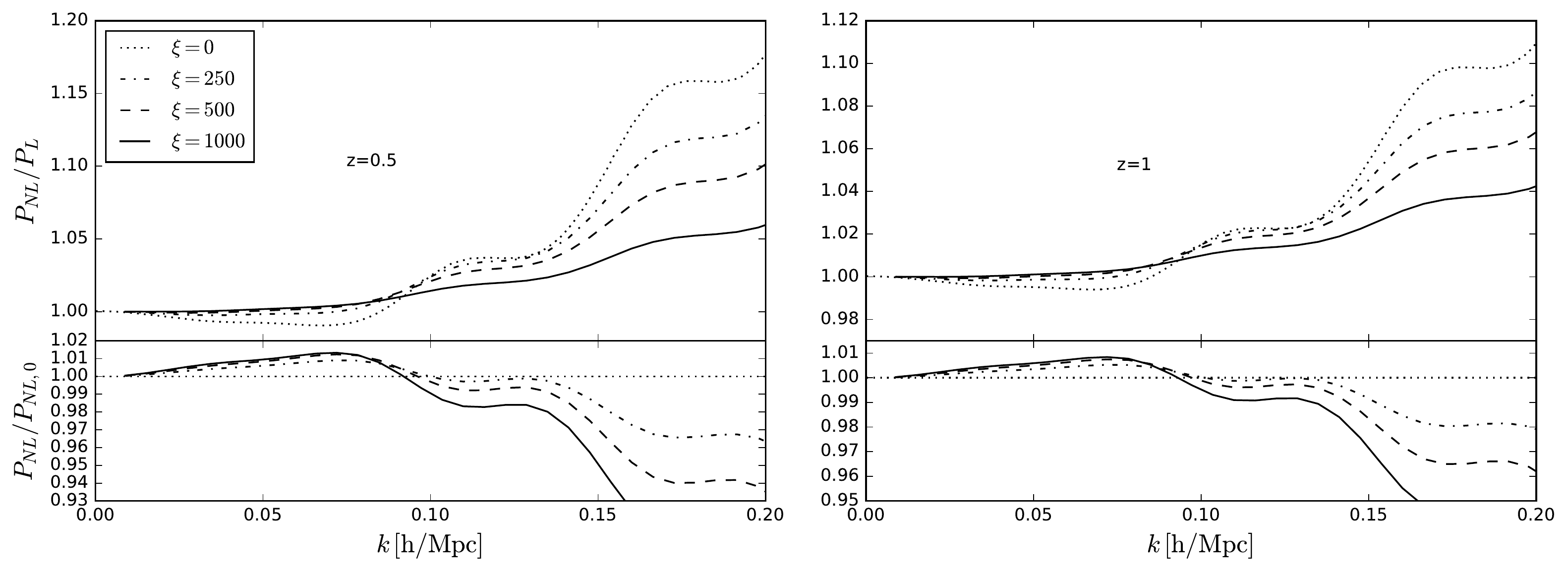}}
\caption[CONVERGENCE ]{Same as Fig.~\ref{pnl_w09} for CPL.}
\label{pnl_cpl}
\end{figure}

In this Section we provide a test for a modelling that includes no interaction in the context of stage IV spectroscopic galaxy surveys such as Euclid and DESI: assuming an interaction in the dark sector exists, does a modelling with no interaction result in a biased constraint on growth? This will serve two purposes. First to indicate at what level of interaction does one incur a bias by omitting it in theoretical modelling,  and second, at what level of interaction and redshifts can stage IV surveys offer detectable signals of such an interaction. Our analysis will be restricted to dark matter and the Dark Scattering models discussed in this paper, and in this sense will only aim to offer an indication for future analyses of more general dark sector interactions.  
\newline 
\newline 
As we are limited in simulation data, with large sample variance and only low $\xi$, we consider the approach applied in \cite{Bose:2017myh}. This proceeds as follows. First, multipole data is produced for a given interaction model using SPT up to some valid $k_{\rm max}$ given by the N-body comparisons in the previous section. Then the covariance matrix for this data is computed as was done for the N-body data using the parameters of an ideal survey. This data is then given a Gaussian scatter using this covariance matrix, providing an easily produced, idealistic, simulated mock data set. We can then perform a likelihood analysis in an attempt to recover the fiducial growth of structure, $f$. 
\newline 
\newline 
For our mock data we consider the $w = -0.9, -1.1$ and CPL models at $z=0.5$ and $z=1$. For these redshifts, the range of validity will be set to $k_{\rm max} = 0.153 \, h$/Mpc and $k_{\rm max} = 0.178 \, h$/Mpc, which are taken as conservative limits based on Figures~\ref{p02wfixz05} - \ref{p02wvarz1}. Fig.~\ref{pxiw09z01} shows the effect of $\xi=100$ peaks at around $\sim 0.5\%$ for the chosen $k_{\rm max}$ for $w = -0.9$, while for the CPL model the effect is $<0.5\%$ for $\xi = 50$. Based on this, data sets will be constructed for $\xi = 250, 500$ and $1000$ as extreme cases. We will also take volumes to reflect realistic surveys, with $V_s=10 \, \mbox{Gpc}^3/h^3$ for $z=0.5$ and $V_s=20 \, \mbox{Gpc}^3/h^3$ for $z=1$. The shot-noise term will be $\bar{n}= 5 \times 10^{-3} \, h^3/\mbox{Mpc}^{3}$ for both redshifts. We summarize all this in Table~\ref{datastats}. The values reflect the survey parameters for DESI \cite{Zhao:2013dza,Aghamousa:2016zmz}. The volume used here is conservatively smaller than the upcoming EUCLID survey, which aims to survey a volume around 3 times as large \cite{Laureijs:2011gra}.
\newline 
\newline 
Fig.~\ref{frecovery1} shows the marginalized best fit growth and $2\sigma$ errors for a parameter inference analysis using the theoretical modelling of Eq.~(\ref{redshiftps}) where $f$ and $\sigma_v$ are treated as the free parameters\footnote{In our analysis we fix the normalization of the power spectrum to the fiducial value.}. This modelling assumes LCDM and so $\xi=0$. This is done against our sets of mock data. The best fit $\sigma_v$ for the analysis are shown in Table~\ref{sumres1}. We see that in general, even for these extreme cases and in the context of high precision data, the interaction signal is weak and/or can be absorbed by $\sigma_v$. The exception is the $w=-1.1$ case, with a large bias on estimated growth from the LCDM modelling showing up at $z=0.5$ in the $\xi=1000$ case. One should note that a smaller $k_{\rm max}$ will produce larger errors on $f$ simply because we are using less information, but will be expected to include the central value at our chosen $k_{\rm max}$. We remind the reader that $k_{\rm max}$ is restricted by the regime that SPT is valid within. At lower redshift SPT begins to break down at larger scales and so we must choose a smaller $k_{\rm max}$, but at this redshift the non-linear signal of interaction is enhanced as structure has had more time to grow. Vice versa, at high redshift, we can choose a higher $k_{\rm max}$, but the non-linear signal will be smaller than at lower redshifts. We comment more on the importance of pushing $k_{\rm max}$ to larger values in the Conclusions. 
\newline 
\newline 
To elucidate the results shown in Fig.~\ref{frecovery1}, we plot the non-linear signals in the matter power spectrum. Fig.~\ref{pnl_w09}, Fig.~\ref{pnl_w11} and Fig.~\ref{pnl_cpl} show the non-linear effects of the chosen interaction  parameters $\xi$. First we comment on the CPL and $w=-0.9$ cases (shown in Fig.~\ref{pnl_w09} and Fig.~\ref{pnl_cpl}). We find that for $z=0.5$ the interaction signal for $\xi=1000$ peaks at $5\%$ at $k=0.15$, where it would become partly degenerate with $\sigma_v$. There is a $\sim 1.5\%$ effect in the large scale regime. The effects are of similar magnitude for $z=1$ and slightly less for the $w=-0.9$ case when compared to CPL. 
\newline 
\newline 
On the other hand, the $w = -1.1$ case shows much stronger enhancements coming from interaction when $\xi$ is very large. Furthermore, these have a very different shape than the $w=-0.9$ or CPL case. The interaction suppresses power over a wide range of scales and is much more prominent at $z=0.5$. This shows why we get the strong bias in the $\xi=0$ model's recovery of growth for data where $\xi=1000$ at $z=0.5$. There is still a bias at $z=1$ but the magnitude of the effect is much smaller here. The plot suggests that $f$ and $\sigma_v$ cannot capture the strong shape dependency of these interactions that results in a bias. Table~\ref{sumres1} shows that even for large changes in $\sigma_v$ this is not possible. This is clear as $\sigma_v$ acts to exponentially damp the power and has no other features. We will discuss this further in the Conclusions. 
\newline
\newline
 Fig.\ref{frecovery2} gives an idea of the correlation between $\sigma_v$ and $f$. It shows the 2D constraints for the $\xi=500$, CPL case, from which it is evident that these two parameters do have some degeneracy. Further it shows that the errors shown in Table.\ref{sumres1} and Fig.\ref{frecovery1} are very close to Gaussian. We have checked that this is true for all cases described in this section. 
\begin{table}[ht]
\caption{Mock Data summary for $\xi =250,500$ and $1000 \, \mbox{bn/Gev}$ and $\bar{n}= 5 \times 10^{-3} \, h^3/\mbox{Mpc}^{3}$ . }
\begin{tabular}{c | c | c | c}
\hline \hline 
$z$ & $ V_s [\mbox{Gev}^3/h^3] $ & $\sigma_v^{\rm fiducial}$ & $k_{\rm max}$ (bins) \\ 
\hline
1 & 20 & $4.5$ & 0.178 (28) \\
0.5 & 10 & $3$ & 0.153 (24)\\
\hline
\end{tabular}
\label{datastats}
\end{table}
\begin{table}[ht]
\caption{Best fit $\sigma_v$ in mock data analysis with $\sigma_v^{\rm fiducial} = 4.5 \, \mbox{Mpc}/h$ ($z=1$) and $\sigma_v^{\rm fiducial} = 3 \, \mbox{Mpc}/h$ ($z=0.5$). The $\chi^2$ of the fit is included in parenthesis.}
\begin{tabular}{c | c | c | c | c}
\hline \hline 
$z$ & $\xi [\mbox{bn/Gev}] $ & $\sigma_v\pm 2\sigma$ $(w=-0.9)$ & $\sigma_v\pm 2\sigma$ $(w=-1.1)$ & $\sigma_v\pm 2\sigma$ (CPL) \\ 
\hline
1 & 250 & $4.35\pm_{0.10}^{0.10}$ (1.78)  & $4.61\pm_{0.09}^{0.09}$(1.74) &$4.44\pm_{0.10}^{0.08}$ (2.99)  \\
1 & 500 & $4.33\pm_{0.11}^{0.10}$ (3.01) & $4.67\pm_{0.09}^{0.08}$(4.42) &$4.39\pm_{0.10}^{0.09}$(2.29) \\
1 & 1000 & $4.35\pm_{0.10}^{0.11}$ (3.07) & $5.56\pm_{0.07}^{0.07}$ (22.67) &$4.44\pm_{0.11}^{0.11}$ (2.55) \\
0.5 & 250 & $2.57\pm_{0.34}^{0.33}$ (1.66) & $3.41\pm_{0.34}^{0.28}$ (2.69) &$2.70\pm_{0.32}^{0.36}$ (2.75) \\
0.5 & 500 & $2.73\pm_{0.31}^{0.28}$ (3.75) & $4.61\pm_{0.20}^{0.21}$ (10.25) &$2.31\pm_{0.29}^{0.31}$ (2.46)\\
0.5 & 1000 &$ 2.81\pm_{0.36}^{0.31}$ (2.53) & $8.21\pm_{0.12}^{0.11}$ (21.79)&$2.74\pm_{0.34}^{0.30}$ (2.54)\\
\hline
\end{tabular}
\label{sumres1}
\end{table}

\newpage
\section{Summary and Conclusions}
This work comes as a first step in preparing and directing analysis pipelines for interesting competitors to the concordance model of cosmology. Specifically, we present the ability of perturbation theory at the 1-loop level in modelling momentum exchange in the dark sector. Momentum exchange between dark energy and dark matter is an interesting alternative to LCDM: it can provide a very good fit to the CMB data and is also able to reconcile discrepancy between CMB and low redshift probes. Furthermore, general parameterized equations of state of DE can still fit the data \cite{Pantazis:2016nky} and will be tested by future galaxy clustering and lensing surveys like Euclid and LSST \cite{Abate:2012za}. 
\newline
\newline
On this note, $w = -1$ and no dark sector interaction are the two assumptions we choose to relax. We consider two evolving and two constant equations of state, with none deviating from a LCDM background history by more than $2.5\%$. We also investigate the signals of momentum exchange and their prominence at different redshifts and when using different DE equations of state. This leads to a first level analysis of whether or not standard perturbative templates can do without modelling the interaction in the context of upcoming spectroscopic surveys.
\newline 
\newline 
We concentrate on the phenomenological Dark Scattering model \cite{Simpson:2010vh,Baldi:2016zom} that benefits from simplicity as well as available N-body data to compare the perturbative approach with. It also has a well defined LCDM limit. We begin by comparing the real space spectra for four different equations of state, two constant and two evolving, and different levels of interaction. The perturbative treatment is shown to do very well in comparison to the N-body data, and is capable of modelling the interaction signal at linear and quasi non-linear scales. The linear effect of the interaction is an overall suppression/enhancement ($w = -0.9/-1.1$) of power, which is degenerate with the power spectrum amplitude. The interaction's non-linear effect entering the 1-loop terms has a scale dependence but it is found to be at the sub percent level, even at $z=0$ with $\xi<100 \, \mbox{bn/Gev}$. Despite this, the perturbative treatment picks it up well. In the evolving $w$ model case, this effect is only slightly larger.  
\newline 
\newline 
It is found that for the constant $w$ cases the non-linear effects of the interaction are split in two regimes. For the $w=-0.9$ case, we get a small enhancement of power at large scales and a suppression of power at smaller scales compared to the $\xi=0$ case. This is the opposite for $w = -1.1$. Again these effects are tiny in the SPT validity regime for $\xi=10$, which is the value of the N-body data. We find that the perturbative predictions for these effects grow with $\xi$. It is also found that the CPL and HYP models have similar non-linear effects that are in turn similar in shape to the $w = -0.9$ case. This was also found to be true in \cite{Baldi:2016zom}, despite these models having very different evolutions as well as drag term, $A$. Effectively, $\xi >0$ in these 3 cases results in smaller loop contributions to the linear power spectrum at $k\leq0.2 \, h$/Mpc for $z\leq 1$. The $w =-1.1$ case results in larger loop contributions. For the $\xi$ considered, these effects are negligible where the loop terms are small ($k<0.1 \, h$/Mpc) but become noticeable at smaller scales. One can see this in Fig.~8 of \cite{Baldi:2016zom} where we see a maximal enhancement ($\xi = -1.1$) and suppression ($\xi = -0.9$, CPL2, HYP1) in power compared to LCDM at around $k=0.4 \, h$/Mpc. At smaller scales the effect is reversed and we get dramatic changes in power compared to LCDM.  
\newline 
\newline 
Next we compare the redshift space monopole and quadrupole at $z=0.5$ and $z=1$, which are observationally relevant for upcoming surveys. We find that the TNS model for the redshift space spectrum does very well in modelling the multipoles. The added flexibility of the free, small scale damping parameter, $\sigma_v$, allows us to push to larger $k$ and captures the overall non-linear effect of $w \neq -1$ and $\xi>0$. These work to suppress multipole power for $w =-1.1$ and enhance it for the $w = -0.9$, CPL and HYP models. There is also shape dependency, which is more prominent in the evolving $w$ models at the relevant redshifts and $\xi$ parameter values. To disentangle the effect of $\xi>0$ from $w \neq -1$ and thus quantify the quality of the modelling, one could perform a likelihood analysis on the data.
\newline 
\newline 
The fact that we are limited in realisations, box size, resolution and interaction strength makes an accurate, robust statistical analysis difficult. To get around this, we create mock data sets using perturbation theory with a covariance matrix constructed using linear theory and future survey parameters. Specifically, we use $V_s=10 \, \mbox{Gpc}^3/h^3$ for $z=0.5$ and $V_s=20 \, \mbox{Gpc}^3/h^3$ for $z=1$. We also include a shot noise term using a dark matter number density $\bar{n}= 5 \times 10^{-3} \, h^3/\mbox{Mpc}^{3}$. The former volume and number density are comparable to those of the BOSS CMASS sample \cite{Alam:2015mbd} or DESI's Luminous Red Galaxy (LRG) target sample \cite{Zhao:2013dza,Aghamousa:2016zmz}. The $z=1$ volume is DESI and Euclid-like \cite{Laureijs:2011gra}. This is done for large values of the interaction parameter $\xi = 250, 500$, and $1000$. The perturbative framework is then applied to this data with $\xi=0$ and keeping $\sigma_v$ and $f$ free. We find that for $w = -0.9$ and CPL, this approach does very well in recovering the fiducial growth of the data, only struggling slightly at $\xi = 500$. This may be explained in terms of these model's prominent effects kicking in at small scales. We find that in this regime, $\sigma_v$ is degenerate with the effects to a sufficient extent. This is supported by Table~\ref{sumres1} which shows smaller values of $\sigma_v$ in these models than the fiducial one indicating they reduce the damping to account for the suppression of non-linearities by $\xi$. 
 \newline 
 \newline 
 On the other hand, the $w = -1.1$ case shows very different results with the $\xi=0$ modelling struggling to recover the fiducial growth for $\xi = 500$ and failing completely for $\xi=1000$, especially at $z=0.5$. This is supported by Fig.~\ref{pnl_w11} which shows very prominent suppression of power over a large range of scales at $z=0.5$. Of course we have not included the full set of galaxy clustering analysis parameters such as tracer bias, or Alcock-Paczynski parameters, although the strong shape dependency suggests that a lot of freedom will be needed to introduce a degeneracy with the effect. Another caveat is that we have not validated the SPT range of validity for such strong couplings, although the deviation from $\xi=0$ is still very prominent at larger scales ($k\leq 0.1h$/Mpc) where SPT is known to do very well even at lower redshifts. In any case, we await future simulation data to perform the full analysis. 
 \newline 
 \newline 
At this stage we comment that interaction with $w< -1$ will be an interesting case for future analyses. This so called phantom DE implies that the energy density of DE is increasing with time (we refer the interested reader to \cite{Caldwell:2003vq} for a discussion on this topic). Phantom DE has already been tested with Planck, and some data combinations tend to pull $w$ into the phantom regime \cite{Planck:2015xua}. Furthermore, we note that the effect of $\xi$ on the linear perturbations acts efficiently to suppress/enhance the amplitude of the power spectrum. For example, for $z=0$, $\xi = 10$ and $w= -0.9$ we get a $1.5\%$ suppression of linear growth over $\xi=0$. The non-linear effect in SPT's validity regime for these parameters is below sub-percent (see bottom panel of Fig.~\ref{pxiw09z01}). Large values of $\xi$ have a large effect on the linear growth; for $\xi=500$ the effect on the linear growth is $ \sim 75\%$. 
\newline
\newline
By moving to smaller scales we expect the signal of momentum transfer to become stronger as indicated by Fig.~8 of \cite{Baldi:2016zom}, with the maximum signature at $k\sim0.4 \, h$/Mpc. These scales are claimed to be modelled well, especially at higher $z$, by the EFToLSS \cite{Baumann:2010tm,Carrasco:2012cv,Lewandowski:2015ziq,Perko:2016puo,delaBella:2017qjy}. This framework would then provide an excellent means of probing the dark sector. We leave this to a future work. In a forthcoming work, we will also extend this framework and analysis to more general momentum exchange models \cite{Pourtsidou:2016ico,Linton:2017ged}. These models use a Lagrangian approach that is physical and self-consistent, and this has important implications for the behaviour of $w$, which is directly affected by the interaction via the modified kinetic term of the DE field \cite{Pourtsidou:2016ico,Linton:2017ged}. We also aim to use efficient simulation approaches such as PICOLA \cite{Howlett:2015hfa} to test the modelling. Such approaches have already been extended to general modified gravity theories \cite{Winther:2017jof} and quintessence-type theories.  

\section*{Acknowledgments}
\noindent The authors would like to thank the anonymous referee for their suggestions and comments. BB is supported by the University of Portsmouth. 
MB acknowledges support from the Italian Ministry for Education, University and Research (MIUR)
through the SIR individual grant SIMCODE, project number RBSI14P4IH, and from the grant MIUR PRIN 2015
``Cosmology and Fundamental Physics: illuminating the Dark
Universe with Euclid".
AP's work for this project was partly supported by a Dennis Sciama Fellowship at the University of Portsmouth. The authors would like to thank Rob Crittenden, Mark Linton, and Kazuya Koyama for very useful discussions.
\appendix
\renewcommand{\bibname}{References}
\bibliography{mybib}{}

\begin{thebibliography}{75}%
\makeatletter
\providecommand \@ifxundefined [1]{%
 \@ifx{#1\undefined}
}%
\providecommand \@ifnum [1]{%
 \ifnum #1\expandafter \@firstoftwo
 \else \expandafter \@secondoftwo
 \fi
}%
\providecommand \@ifx [1]{%
 \ifx #1\expandafter \@firstoftwo
 \else \expandafter \@secondoftwo
 \fi
}%
\providecommand \natexlab [1]{#1}%
\providecommand \enquote  [1]{``#1''}%
\providecommand \bibnamefont  [1]{#1}%
\providecommand \bibfnamefont [1]{#1}%
\providecommand \citenamefont [1]{#1}%
\providecommand \href@noop [0]{\@secondoftwo}%
\providecommand \href [0]{\begingroup \@sanitize@url \@href}%
\providecommand \@href[1]{\@@startlink{#1}\@@href}%
\providecommand \@@href[1]{\endgroup#1\@@endlink}%
\providecommand \@sanitize@url [0]{\catcode `\\12\catcode `\$12\catcode
  `\&12\catcode `\#12\catcode `\^12\catcode `\_12\catcode `\%12\relax}%
\providecommand \@@startlink[1]{}%
\providecommand \@@endlink[0]{}%
\providecommand \url  [0]{\begingroup\@sanitize@url \@url }%
\providecommand \@url [1]{\endgroup\@href {#1}{\urlprefix }}%
\providecommand \urlprefix  [0]{URL }%
\providecommand \Eprint [0]{\href }%
\providecommand \doibase [0]{http://dx.doi.org/}%
\providecommand \selectlanguage [0]{\@gobble}%
\providecommand \bibinfo  [0]{\@secondoftwo}%
\providecommand \bibfield  [0]{\@secondoftwo}%
\providecommand \translation [1]{[#1]}%
\providecommand \BibitemOpen [0]{}%
\providecommand \bibitemStop [0]{}%
\providecommand \bibitemNoStop [0]{.\EOS\space}%
\providecommand \EOS [0]{\spacefactor3000\relax}%
\providecommand \BibitemShut  [1]{\csname bibitem#1\endcsname}%
\let\auto@bib@innerbib\@empty
\bibitem [{\citenamefont {Ade}\ \emph {et~al.}(2016)\citenamefont {Ade} \emph
  {et~al.}}]{Planck:2015xua}%
  \BibitemOpen
  \bibfield  {author} {\bibinfo {author} {\bibfnamefont {P.~A.~R.}\
  \bibnamefont {Ade}} \emph {et~al.} (\bibinfo {collaboration} {Planck}),\
  }\href {\doibase 10.1051/0004-6361/201525830} {\bibfield  {journal} {\bibinfo
   {journal} {Astron. Astrophys.}\ }\textbf {\bibinfo {volume} {594}},\
  \bibinfo {pages} {A13} (\bibinfo {year} {2016})},\ \Eprint
  {http://arxiv.org/abs/1502.01589} {arXiv:1502.01589 [astro-ph.CO]}
  \BibitemShut {NoStop}%
\bibitem [{\citenamefont {Anderson}\ \emph {et~al.}(2014)\citenamefont
  {Anderson} \emph {et~al.}}]{Anderson:2013zyy}%
  \BibitemOpen
  \bibfield  {author} {\bibinfo {author} {\bibfnamefont {L.}~\bibnamefont
  {Anderson}} \emph {et~al.} (\bibinfo {collaboration} {BOSS}),\ }\href
  {\doibase 10.1093/mnras/stu523} {\bibfield  {journal} {\bibinfo  {journal}
  {Mon. Not. Roy. Astron. Soc.}\ }\textbf {\bibinfo {volume} {441}},\ \bibinfo
  {pages} {24} (\bibinfo {year} {2014})},\ \Eprint
  {http://arxiv.org/abs/1312.4877} {arXiv:1312.4877 [astro-ph.CO]} \BibitemShut
  {NoStop}%
\bibitem [{\citenamefont {Riess}\ \emph {et~al.}(2009)\citenamefont {Riess}
  \emph {et~al.}}]{Riess:2009pu}%
  \BibitemOpen
  \bibfield  {author} {\bibinfo {author} {\bibfnamefont {A.~G.}\ \bibnamefont
  {Riess}} \emph {et~al.},\ }\href {\doibase 10.1088/0004-637X/699/1/539}
  {\bibfield  {journal} {\bibinfo  {journal} {Astrophys. J.}\ }\textbf
  {\bibinfo {volume} {699}},\ \bibinfo {pages} {539} (\bibinfo {year}
  {2009})},\ \Eprint {http://arxiv.org/abs/0905.0695} {arXiv:0905.0695
  [astro-ph.CO]} \BibitemShut {NoStop}%
\bibitem [{\citenamefont {Lampeitl}\ \emph {et~al.}(2009)\citenamefont
  {Lampeitl} \emph {et~al.}}]{Lampeitl:2009jq}%
  \BibitemOpen
  \bibfield  {author} {\bibinfo {author} {\bibfnamefont {H.}~\bibnamefont
  {Lampeitl}} \emph {et~al.},\ }\href {\doibase
  10.1111/j.1365-2966.2009.15851.x} {\bibfield  {journal} {\bibinfo  {journal}
  {Mon. Not. Roy. Astron. Soc.}\ }\textbf {\bibinfo {volume} {401}},\ \bibinfo
  {pages} {2331} (\bibinfo {year} {2009})},\ \Eprint
  {http://arxiv.org/abs/0910.2193} {arXiv:0910.2193 [astro-ph.CO]} \BibitemShut
  {NoStop}%
\bibitem [{\citenamefont {Weinberg}(1989)}]{Weinberg:1988cp}%
  \BibitemOpen
  \bibfield  {author} {\bibinfo {author} {\bibfnamefont {S.}~\bibnamefont
  {Weinberg}},\ }\href {\doibase 10.1103/RevModPhys.61.1} {\bibfield  {journal}
  {\bibinfo  {journal} {Rev.Mod.Phys.}\ }\textbf {\bibinfo {volume} {61}},\
  \bibinfo {pages} {1} (\bibinfo {year} {1989})}\BibitemShut {NoStop}%
\bibitem [{\citenamefont {Martin}(2012)}]{Martin:2012bt}%
  \BibitemOpen
  \bibfield  {author} {\bibinfo {author} {\bibfnamefont {J.}~\bibnamefont
  {Martin}},\ }\href {\doibase 10.1016/j.crhy.2012.04.008} {\bibfield
  {journal} {\bibinfo  {journal} {Comptes Rendus Physique}\ }\textbf {\bibinfo
  {volume} {13}},\ \bibinfo {pages} {566} (\bibinfo {year} {2012})},\ \Eprint
  {http://arxiv.org/abs/1205.3365} {arXiv:1205.3365 [astro-ph.CO]} \BibitemShut
  {NoStop}%
\bibitem [{\citenamefont {Copeland}\ \emph {et~al.}(2006)\citenamefont
  {Copeland}, \citenamefont {Sami},\ and\ \citenamefont
  {Tsujikawa}}]{Copeland:2006wr}%
  \BibitemOpen
  \bibfield  {author} {\bibinfo {author} {\bibfnamefont {E.~J.}\ \bibnamefont
  {Copeland}}, \bibinfo {author} {\bibfnamefont {M.}~\bibnamefont {Sami}}, \
  and\ \bibinfo {author} {\bibfnamefont {S.}~\bibnamefont {Tsujikawa}},\ }\href
  {\doibase 10.1142/S021827180600942X} {\bibfield  {journal} {\bibinfo
  {journal} {Int.J.Mod.Phys.}\ }\textbf {\bibinfo {volume} {D15}},\ \bibinfo
  {pages} {1753} (\bibinfo {year} {2006})},\ \Eprint
  {http://arxiv.org/abs/hep-th/0603057} {arXiv:hep-th/0603057 [hep-th]}
  \BibitemShut {NoStop}%
\bibitem [{\citenamefont {Clifton}\ \emph {et~al.}(2012)\citenamefont
  {Clifton}, \citenamefont {Ferreira}, \citenamefont {Padilla},\ and\
  \citenamefont {Skordis}}]{Clifton:2011jh}%
  \BibitemOpen
  \bibfield  {author} {\bibinfo {author} {\bibfnamefont {T.}~\bibnamefont
  {Clifton}}, \bibinfo {author} {\bibfnamefont {P.~G.}\ \bibnamefont
  {Ferreira}}, \bibinfo {author} {\bibfnamefont {A.}~\bibnamefont {Padilla}}, \
  and\ \bibinfo {author} {\bibfnamefont {C.}~\bibnamefont {Skordis}},\ }\href
  {\doibase 10.1016/j.physrep.2012.01.001} {\bibfield  {journal} {\bibinfo
  {journal} {Phys.Rept.}\ }\textbf {\bibinfo {volume} {513}},\ \bibinfo {pages}
  {1} (\bibinfo {year} {2012})},\ \Eprint {http://arxiv.org/abs/1106.2476}
  {arXiv:1106.2476 [astro-ph.CO]} \BibitemShut {NoStop}%
\bibitem [{\citenamefont {Vikhlinin}\ \emph {et~al.}(2009)\citenamefont
  {Vikhlinin} \emph {et~al.}}]{Vikhlinin:2008ym}%
  \BibitemOpen
  \bibfield  {author} {\bibinfo {author} {\bibfnamefont {A.}~\bibnamefont
  {Vikhlinin}} \emph {et~al.},\ }\href {\doibase 10.1088/0004-637X/692/2/1060}
  {\bibfield  {journal} {\bibinfo  {journal} {Astrophys. J.}\ }\textbf
  {\bibinfo {volume} {692}},\ \bibinfo {pages} {1060} (\bibinfo {year}
  {2009})},\ \Eprint {http://arxiv.org/abs/0812.2720} {arXiv:0812.2720
  [astro-ph]} \BibitemShut {NoStop}%
\bibitem [{\citenamefont {de~Haan}\ \emph {et~al.}(2016)\citenamefont {de~Haan}
  \emph {et~al.}}]{deHaan:2016qvy}%
  \BibitemOpen
  \bibfield  {author} {\bibinfo {author} {\bibfnamefont {T.}~\bibnamefont
  {de~Haan}} \emph {et~al.} (\bibinfo {collaboration} {SPT}),\ }\href {\doibase
  10.3847/0004-637X/832/1/95} {\bibfield  {journal} {\bibinfo  {journal}
  {Astrophys. J.}\ }\textbf {\bibinfo {volume} {832}},\ \bibinfo {pages} {95}
  (\bibinfo {year} {2016})},\ \Eprint {http://arxiv.org/abs/1603.06522}
  {arXiv:1603.06522 [astro-ph.CO]} \BibitemShut {NoStop}%
\bibitem [{\citenamefont {Heymans}\ \emph {et~al.}(2013)\citenamefont {Heymans}
  \emph {et~al.}}]{Heymans:2013fya}%
  \BibitemOpen
  \bibfield  {author} {\bibinfo {author} {\bibfnamefont {C.}~\bibnamefont
  {Heymans}} \emph {et~al.},\ }\href {\doibase 10.1093/mnras/stt601} {\bibfield
   {journal} {\bibinfo  {journal} {Mon. Not. Roy. Astron. Soc.}\ }\textbf
  {\bibinfo {volume} {432}},\ \bibinfo {pages} {2433} (\bibinfo {year}
  {2013})},\ \Eprint {http://arxiv.org/abs/1303.1808} {arXiv:1303.1808
  [astro-ph.CO]} \BibitemShut {NoStop}%
\bibitem [{\citenamefont {Abbott}\ \emph {et~al.}(2017)\citenamefont {Abbott}
  \emph {et~al.}}]{Abbott:2017wau}%
  \BibitemOpen
  \bibfield  {author} {\bibinfo {author} {\bibfnamefont {T.~M.~C.}\
  \bibnamefont {Abbott}} \emph {et~al.} (\bibinfo {collaboration} {DES}),\
  }\href@noop {} {\  (\bibinfo {year} {2017})},\ \Eprint
  {http://arxiv.org/abs/1708.01530} {arXiv:1708.01530 [astro-ph.CO]}
  \BibitemShut {NoStop}%
\bibitem [{\citenamefont {Macaulay}\ \emph {et~al.}(2013)\citenamefont
  {Macaulay}, \citenamefont {Wehus},\ and\ \citenamefont
  {Eriksen}}]{Macaulay:2013swa}%
  \BibitemOpen
  \bibfield  {author} {\bibinfo {author} {\bibfnamefont {E.}~\bibnamefont
  {Macaulay}}, \bibinfo {author} {\bibfnamefont {I.~K.}\ \bibnamefont {Wehus}},
  \ and\ \bibinfo {author} {\bibfnamefont {H.~K.}\ \bibnamefont {Eriksen}},\
  }\href {\doibase 10.1103/PhysRevLett.111.161301} {\bibfield  {journal}
  {\bibinfo  {journal} {Phys. Rev. Lett.}\ }\textbf {\bibinfo {volume} {111}},\
  \bibinfo {pages} {161301} (\bibinfo {year} {2013})},\ \Eprint
  {http://arxiv.org/abs/1303.6583} {arXiv:1303.6583 [astro-ph.CO]} \BibitemShut
  {NoStop}%
\bibitem [{\citenamefont {Blake}\ \emph {et~al.}(2011)\citenamefont {Blake}
  \emph {et~al.}}]{Blake:2011rj}%
  \BibitemOpen
  \bibfield  {author} {\bibinfo {author} {\bibfnamefont {C.}~\bibnamefont
  {Blake}} \emph {et~al.},\ }\href {\doibase 10.1111/j.1365-2966.2011.18903.x}
  {\bibfield  {journal} {\bibinfo  {journal} {Mon. Not. Roy. Astron. Soc.}\
  }\textbf {\bibinfo {volume} {415}},\ \bibinfo {pages} {2876} (\bibinfo {year}
  {2011})},\ \Eprint {http://arxiv.org/abs/1104.2948} {arXiv:1104.2948
  [astro-ph.CO]} \BibitemShut {NoStop}%
\bibitem [{\citenamefont {Reid}\ \emph {et~al.}(2012)\citenamefont {Reid} \emph
  {et~al.}}]{Reid:2012sw}%
  \BibitemOpen
  \bibfield  {author} {\bibinfo {author} {\bibfnamefont {B.~A.}\ \bibnamefont
  {Reid}} \emph {et~al.},\ }\href {\doibase 10.1111/j.1365-2966.2012.21779.x}
  {\bibfield  {journal} {\bibinfo  {journal} {Mon. Not. Roy. Astron. Soc.}\
  }\textbf {\bibinfo {volume} {426}},\ \bibinfo {pages} {2719} (\bibinfo {year}
  {2012})},\ \Eprint {http://arxiv.org/abs/1203.6641} {arXiv:1203.6641
  [astro-ph.CO]} \BibitemShut {NoStop}%
\bibitem [{\citenamefont {Beutler}\ \emph {et~al.}(2014)\citenamefont {Beutler}
  \emph {et~al.}}]{Beutler:2013yhm}%
  \BibitemOpen
  \bibfield  {author} {\bibinfo {author} {\bibfnamefont {F.}~\bibnamefont
  {Beutler}} \emph {et~al.} (\bibinfo {collaboration} {BOSS}),\ }\href
  {\doibase 10.1093/mnras/stu1051} {\bibfield  {journal} {\bibinfo  {journal}
  {Mon. Not. Roy. Astron. Soc.}\ }\textbf {\bibinfo {volume} {443}},\ \bibinfo
  {pages} {1065} (\bibinfo {year} {2014})},\ \Eprint
  {http://arxiv.org/abs/1312.4611} {arXiv:1312.4611 [astro-ph.CO]} \BibitemShut
  {NoStop}%
\bibitem [{\citenamefont {Gil-Marin}\ \emph {et~al.}(2016)\citenamefont
  {Gil-Marin} \emph {et~al.}}]{Gil-Marin:2015sqa}%
  \BibitemOpen
  \bibfield  {author} {\bibinfo {author} {\bibfnamefont {H.}~\bibnamefont
  {Gil-Marin}} \emph {et~al.},\ }\href {\doibase 10.1093/mnras/stw1096}
  {\bibfield  {journal} {\bibinfo  {journal} {Mon. Not. Roy. Astron. Soc.}\
  }\textbf {\bibinfo {volume} {460}},\ \bibinfo {pages} {4188} (\bibinfo {year}
  {2016})},\ \Eprint {http://arxiv.org/abs/1509.06386} {arXiv:1509.06386
  [astro-ph.CO]} \BibitemShut {NoStop}%
\bibitem [{\citenamefont {Simpson}\ \emph {et~al.}(2016)\citenamefont
  {Simpson}, \citenamefont {Blake}, \citenamefont {Peacock}, \citenamefont
  {Baldry}, \citenamefont {Bland-Hawthorn}, \citenamefont {Heavens},
  \citenamefont {Heymans}, \citenamefont {Loveday},\ and\ \citenamefont
  {Norberg}}]{Simpson:2015yfa}%
  \BibitemOpen
  \bibfield  {author} {\bibinfo {author} {\bibfnamefont {F.}~\bibnamefont
  {Simpson}}, \bibinfo {author} {\bibfnamefont {C.}~\bibnamefont {Blake}},
  \bibinfo {author} {\bibfnamefont {J.~A.}\ \bibnamefont {Peacock}}, \bibinfo
  {author} {\bibfnamefont {I.}~\bibnamefont {Baldry}}, \bibinfo {author}
  {\bibfnamefont {J.}~\bibnamefont {Bland-Hawthorn}}, \bibinfo {author}
  {\bibfnamefont {A.}~\bibnamefont {Heavens}}, \bibinfo {author} {\bibfnamefont
  {C.}~\bibnamefont {Heymans}}, \bibinfo {author} {\bibfnamefont
  {J.}~\bibnamefont {Loveday}}, \ and\ \bibinfo {author} {\bibfnamefont
  {P.}~\bibnamefont {Norberg}},\ }\href {\doibase 10.1103/PhysRevD.93.023525}
  {\bibfield  {journal} {\bibinfo  {journal} {Phys. Rev.}\ }\textbf {\bibinfo
  {volume} {D93}},\ \bibinfo {pages} {023525} (\bibinfo {year} {2016})},\
  \Eprint {http://arxiv.org/abs/1505.03865} {arXiv:1505.03865 [astro-ph.CO]}
  \BibitemShut {NoStop}%
\bibitem [{\citenamefont {Simpson}(2010)}]{Simpson:2010vh}%
  \BibitemOpen
  \bibfield  {author} {\bibinfo {author} {\bibfnamefont {F.}~\bibnamefont
  {Simpson}},\ }\href {\doibase 10.1103/PhysRevD.82.083505} {\bibfield
  {journal} {\bibinfo  {journal} {Phys. Rev.}\ }\textbf {\bibinfo {volume}
  {D82}},\ \bibinfo {pages} {083505} (\bibinfo {year} {2010})},\ \Eprint
  {http://arxiv.org/abs/1007.1034} {arXiv:1007.1034 [astro-ph.CO]} \BibitemShut
  {NoStop}%
\bibitem [{\citenamefont {Lesgourgues}\ \emph {et~al.}(2016)\citenamefont
  {Lesgourgues}, \citenamefont {Marques-Tavares},\ and\ \citenamefont
  {Schmaltz}}]{Lesgourgues:2015wza}%
  \BibitemOpen
  \bibfield  {author} {\bibinfo {author} {\bibfnamefont {J.}~\bibnamefont
  {Lesgourgues}}, \bibinfo {author} {\bibfnamefont {G.}~\bibnamefont
  {Marques-Tavares}}, \ and\ \bibinfo {author} {\bibfnamefont {M.}~\bibnamefont
  {Schmaltz}},\ }\href {\doibase 10.1088/1475-7516/2016/02/037} {\bibfield
  {journal} {\bibinfo  {journal} {JCAP}\ }\textbf {\bibinfo {volume} {1602}},\
  \bibinfo {pages} {037} (\bibinfo {year} {2016})},\ \Eprint
  {http://arxiv.org/abs/1507.04351} {arXiv:1507.04351 [astro-ph.CO]}
  \BibitemShut {NoStop}%
\bibitem [{\citenamefont {Pourtsidou}\ and\ \citenamefont
  {Tram}(2016)}]{Pourtsidou:2016ico}%
  \BibitemOpen
  \bibfield  {author} {\bibinfo {author} {\bibfnamefont {A.}~\bibnamefont
  {Pourtsidou}}\ and\ \bibinfo {author} {\bibfnamefont {T.}~\bibnamefont
  {Tram}},\ }\href {\doibase 10.1103/PhysRevD.94.043518} {\bibfield  {journal}
  {\bibinfo  {journal} {Phys. Rev.}\ }\textbf {\bibinfo {volume} {D94}},\
  \bibinfo {pages} {043518} (\bibinfo {year} {2016})},\ \Eprint
  {http://arxiv.org/abs/1604.04222} {arXiv:1604.04222 [astro-ph.CO]}
  \BibitemShut {NoStop}%
\bibitem [{\citenamefont {Baldi}\ and\ \citenamefont
  {Simpson}(2017)}]{Baldi:2016zom}%
  \BibitemOpen
  \bibfield  {author} {\bibinfo {author} {\bibfnamefont {M.}~\bibnamefont
  {Baldi}}\ and\ \bibinfo {author} {\bibfnamefont {F.}~\bibnamefont
  {Simpson}},\ }\href {\doibase 10.1093/mnras/stw2702} {\bibfield  {journal}
  {\bibinfo  {journal} {Mon. Not. Roy. Astron. Soc.}\ }\textbf {\bibinfo
  {volume} {465}},\ \bibinfo {pages} {653} (\bibinfo {year} {2017})},\ \Eprint
  {http://arxiv.org/abs/1605.05623} {arXiv:1605.05623 [astro-ph.CO]}
  \BibitemShut {NoStop}%
\bibitem [{\citenamefont {Buen-Abad}\ \emph {et~al.}(2017)\citenamefont
  {Buen-Abad}, \citenamefont {Schmaltz}, \citenamefont {Lesgourgues},\ and\
  \citenamefont {Brinckmann}}]{Buen-Abad:2017gxg}%
  \BibitemOpen
  \bibfield  {author} {\bibinfo {author} {\bibfnamefont {M.~A.}\ \bibnamefont
  {Buen-Abad}}, \bibinfo {author} {\bibfnamefont {M.}~\bibnamefont {Schmaltz}},
  \bibinfo {author} {\bibfnamefont {J.}~\bibnamefont {Lesgourgues}}, \ and\
  \bibinfo {author} {\bibfnamefont {T.}~\bibnamefont {Brinckmann}},\
  }\href@noop {} {\  (\bibinfo {year} {2017})},\ \Eprint
  {http://arxiv.org/abs/1708.09406} {arXiv:1708.09406 [astro-ph.CO]}
  \BibitemShut {NoStop}%
\bibitem [{\citenamefont {Linton}\ \emph {et~al.}(2017)\citenamefont {Linton},
  \citenamefont {Pourtsidou}, \citenamefont {Crittenden},\ and\ \citenamefont
  {Maartens}}]{Linton:2017ged}%
  \BibitemOpen
  \bibfield  {author} {\bibinfo {author} {\bibfnamefont {M.~S.}\ \bibnamefont
  {Linton}}, \bibinfo {author} {\bibfnamefont {A.}~\bibnamefont {Pourtsidou}},
  \bibinfo {author} {\bibfnamefont {R.}~\bibnamefont {Crittenden}}, \ and\
  \bibinfo {author} {\bibfnamefont {R.}~\bibnamefont {Maartens}},\ }\href@noop
  {} {\  (\bibinfo {year} {2017})},\ \Eprint {http://arxiv.org/abs/1711.05196}
  {arXiv:1711.05196 [astro-ph.CO]} \BibitemShut {NoStop}%
\bibitem [{\citenamefont {Pourtsidou}\ \emph {et~al.}(2013)\citenamefont
  {Pourtsidou}, \citenamefont {Skordis},\ and\ \citenamefont
  {Copeland}}]{Pourtsidou:2013nha}%
  \BibitemOpen
  \bibfield  {author} {\bibinfo {author} {\bibfnamefont {A.}~\bibnamefont
  {Pourtsidou}}, \bibinfo {author} {\bibfnamefont {C.}~\bibnamefont {Skordis}},
  \ and\ \bibinfo {author} {\bibfnamefont {E.~J.}\ \bibnamefont {Copeland}},\
  }\href {\doibase 10.1103/PhysRevD.88.083505} {\bibfield  {journal} {\bibinfo
  {journal} {Phys. Rev.}\ }\textbf {\bibinfo {volume} {D88}},\ \bibinfo {pages}
  {083505} (\bibinfo {year} {2013})},\ \Eprint {http://arxiv.org/abs/1307.0458}
  {arXiv:1307.0458 [astro-ph.CO]} \BibitemShut {NoStop}%
\bibitem [{\citenamefont {Kaiser}(1987)}]{Kaiser:1987qv}%
  \BibitemOpen
  \bibfield  {author} {\bibinfo {author} {\bibfnamefont {N.}~\bibnamefont
  {Kaiser}},\ }\href@noop {} {\bibfield  {journal} {\bibinfo  {journal} {Mon.
  Not. Roy. Astron. Soc.}\ }\textbf {\bibinfo {volume} {227}},\ \bibinfo
  {pages} {1} (\bibinfo {year} {1987})}\BibitemShut {NoStop}%
\bibitem [{\citenamefont {Baldi}\ and\ \citenamefont
  {Simpson}(2015)}]{Baldi:2014ica}%
  \BibitemOpen
  \bibfield  {author} {\bibinfo {author} {\bibfnamefont {M.}~\bibnamefont
  {Baldi}}\ and\ \bibinfo {author} {\bibfnamefont {F.}~\bibnamefont
  {Simpson}},\ }\href {\doibase 10.1093/mnras/stv405} {\bibfield  {journal}
  {\bibinfo  {journal} {Mon. Not. Roy. Astron. Soc.}\ }\textbf {\bibinfo
  {volume} {449}},\ \bibinfo {pages} {2239} (\bibinfo {year} {2015})},\ \Eprint
  {http://arxiv.org/abs/1412.1080} {arXiv:1412.1080 [astro-ph.CO]} \BibitemShut
  {NoStop}%
\bibitem [{\citenamefont {Bernardeau}\ \emph {et~al.}(2002)\citenamefont
  {Bernardeau}, \citenamefont {Colombi}, \citenamefont {Gaztanaga},\ and\
  \citenamefont {Scoccimarro}}]{Bernardeau:2001qr}%
  \BibitemOpen
  \bibfield  {author} {\bibinfo {author} {\bibfnamefont {F.}~\bibnamefont
  {Bernardeau}}, \bibinfo {author} {\bibfnamefont {S.}~\bibnamefont {Colombi}},
  \bibinfo {author} {\bibfnamefont {E.}~\bibnamefont {Gaztanaga}}, \ and\
  \bibinfo {author} {\bibfnamefont {R.}~\bibnamefont {Scoccimarro}},\ }\href
  {\doibase 10.1016/S0370-1573(02)00135-7} {\bibfield  {journal} {\bibinfo
  {journal} {Phys. Rept.}\ }\textbf {\bibinfo {volume} {367}},\ \bibinfo
  {pages} {1} (\bibinfo {year} {2002})},\ \Eprint
  {http://arxiv.org/abs/astro-ph/0112551} {arXiv:astro-ph/0112551 [astro-ph]}
  \BibitemShut {NoStop}%
\bibitem [{\citenamefont {Crocce}\ \emph {et~al.}(2012)\citenamefont {Crocce},
  \citenamefont {Scoccimarro},\ and\ \citenamefont
  {Bernardeau}}]{Crocce:2012fa}%
  \BibitemOpen
  \bibfield  {author} {\bibinfo {author} {\bibfnamefont {M.}~\bibnamefont
  {Crocce}}, \bibinfo {author} {\bibfnamefont {R.}~\bibnamefont {Scoccimarro}},
  \ and\ \bibinfo {author} {\bibfnamefont {F.}~\bibnamefont {Bernardeau}},\
  }\href {\doibase 10.1111/j.1365-2966.2012.22127.x} {\bibfield  {journal}
  {\bibinfo  {journal} {Mon. Not. Roy. Astron. Soc.}\ }\textbf {\bibinfo
  {volume} {427}},\ \bibinfo {pages} {2537} (\bibinfo {year} {2012})},\ \Eprint
  {http://arxiv.org/abs/1207.1465} {arXiv:1207.1465 [astro-ph.CO]} \BibitemShut
  {NoStop}%
\bibitem [{\citenamefont {Taruya}\ and\ \citenamefont
  {Hiramatsu}(2008)}]{Taruya:2007xy}%
  \BibitemOpen
  \bibfield  {author} {\bibinfo {author} {\bibfnamefont {A.}~\bibnamefont
  {Taruya}}\ and\ \bibinfo {author} {\bibfnamefont {T.}~\bibnamefont
  {Hiramatsu}},\ }\href {\doibase 10.1086/526515} {\bibfield  {journal}
  {\bibinfo  {journal} {Astrophys. J.}\ }\textbf {\bibinfo {volume} {674}},\
  \bibinfo {pages} {617} (\bibinfo {year} {2008})},\ \Eprint
  {http://arxiv.org/abs/0708.1367} {arXiv:0708.1367 [astro-ph]} \BibitemShut
  {NoStop}%
\bibitem [{\citenamefont {Pietroni}(2008)}]{Pietroni:2008jx}%
  \BibitemOpen
  \bibfield  {author} {\bibinfo {author} {\bibfnamefont {M.}~\bibnamefont
  {Pietroni}},\ }\href {\doibase 10.1088/1475-7516/2008/10/036} {\bibfield
  {journal} {\bibinfo  {journal} {JCAP}\ }\textbf {\bibinfo {volume} {0810}},\
  \bibinfo {pages} {036} (\bibinfo {year} {2008})},\ \Eprint
  {http://arxiv.org/abs/0806.0971} {arXiv:0806.0971 [astro-ph]} \BibitemShut
  {NoStop}%
\bibitem [{\citenamefont {McDonald}(2007)}]{McDonald:2006hf}%
  \BibitemOpen
  \bibfield  {author} {\bibinfo {author} {\bibfnamefont {P.}~\bibnamefont
  {McDonald}},\ }\href {\doibase 10.1103/PhysRevD.75.043514} {\bibfield
  {journal} {\bibinfo  {journal} {Phys. Rev.}\ }\textbf {\bibinfo {volume}
  {D75}},\ \bibinfo {pages} {043514} (\bibinfo {year} {2007})},\ \Eprint
  {http://arxiv.org/abs/astro-ph/0606028} {arXiv:astro-ph/0606028 [astro-ph]}
  \BibitemShut {NoStop}%
\bibitem [{\citenamefont {Matsubara}(2008)}]{Matsubara:2007wj}%
  \BibitemOpen
  \bibfield  {author} {\bibinfo {author} {\bibfnamefont {T.}~\bibnamefont
  {Matsubara}},\ }\href {\doibase 10.1103/PhysRevD.77.063530} {\bibfield
  {journal} {\bibinfo  {journal} {Phys. Rev.}\ }\textbf {\bibinfo {volume}
  {D77}},\ \bibinfo {pages} {063530} (\bibinfo {year} {2008})},\ \Eprint
  {http://arxiv.org/abs/0711.2521} {arXiv:0711.2521 [astro-ph]} \BibitemShut
  {NoStop}%
\bibitem [{\citenamefont {Bernardeau}\ \emph {et~al.}(2012)\citenamefont
  {Bernardeau}, \citenamefont {Crocce},\ and\ \citenamefont
  {Scoccimarro}}]{Bernardeau:2011dp}%
  \BibitemOpen
  \bibfield  {author} {\bibinfo {author} {\bibfnamefont {F.}~\bibnamefont
  {Bernardeau}}, \bibinfo {author} {\bibfnamefont {M.}~\bibnamefont {Crocce}},
  \ and\ \bibinfo {author} {\bibfnamefont {R.}~\bibnamefont {Scoccimarro}},\
  }\href {\doibase 10.1103/PhysRevD.85.123519} {\bibfield  {journal} {\bibinfo
  {journal} {Phys. Rev.}\ }\textbf {\bibinfo {volume} {D85}},\ \bibinfo {pages}
  {123519} (\bibinfo {year} {2012})},\ \Eprint {http://arxiv.org/abs/1112.3895}
  {arXiv:1112.3895 [astro-ph.CO]} \BibitemShut {NoStop}%
\bibitem [{\citenamefont {Blas}\ \emph {et~al.}(2016)\citenamefont {Blas},
  \citenamefont {Garny}, \citenamefont {Ivanov},\ and\ \citenamefont
  {Sibiryakov}}]{Blas:2015qsi}%
  \BibitemOpen
  \bibfield  {author} {\bibinfo {author} {\bibfnamefont {D.}~\bibnamefont
  {Blas}}, \bibinfo {author} {\bibfnamefont {M.}~\bibnamefont {Garny}},
  \bibinfo {author} {\bibfnamefont {M.~M.}\ \bibnamefont {Ivanov}}, \ and\
  \bibinfo {author} {\bibfnamefont {S.}~\bibnamefont {Sibiryakov}},\ }\href
  {\doibase 10.1088/1475-7516/2016/07/052} {\bibfield  {journal} {\bibinfo
  {journal} {JCAP}\ }\textbf {\bibinfo {volume} {1607}},\ \bibinfo {pages}
  {052} (\bibinfo {year} {2016})},\ \Eprint {http://arxiv.org/abs/1512.05807}
  {arXiv:1512.05807 [astro-ph.CO]} \BibitemShut {NoStop}%
\bibitem [{\citenamefont {Taruya}\ \emph {et~al.}(2010)\citenamefont {Taruya},
  \citenamefont {Nishimichi},\ and\ \citenamefont {Saito}}]{Taruya:2010mx}%
  \BibitemOpen
  \bibfield  {author} {\bibinfo {author} {\bibfnamefont {A.}~\bibnamefont
  {Taruya}}, \bibinfo {author} {\bibfnamefont {T.}~\bibnamefont {Nishimichi}},
  \ and\ \bibinfo {author} {\bibfnamefont {S.}~\bibnamefont {Saito}},\ }\href
  {\doibase 10.1103/PhysRevD.82.063522} {\bibfield  {journal} {\bibinfo
  {journal} {Phys.Rev.}\ }\textbf {\bibinfo {volume} {D82}},\ \bibinfo {pages}
  {063522} (\bibinfo {year} {2010})},\ \Eprint {http://arxiv.org/abs/1006.0699}
  {arXiv:1006.0699 [astro-ph.CO]} \BibitemShut {NoStop}%
\bibitem [{\citenamefont {Nishimichi}\ and\ \citenamefont
  {Taruya}(2011)}]{Nishimichi:2011jm}%
  \BibitemOpen
  \bibfield  {author} {\bibinfo {author} {\bibfnamefont {T.}~\bibnamefont
  {Nishimichi}}\ and\ \bibinfo {author} {\bibfnamefont {A.}~\bibnamefont
  {Taruya}},\ }\href {\doibase 10.1103/PhysRevD.84.043526} {\bibfield
  {journal} {\bibinfo  {journal} {Phys. Rev.}\ }\textbf {\bibinfo {volume}
  {D84}},\ \bibinfo {pages} {043526} (\bibinfo {year} {2011})},\ \Eprint
  {http://arxiv.org/abs/1106.4562} {arXiv:1106.4562 [astro-ph.CO]} \BibitemShut
  {NoStop}%
\bibitem [{\citenamefont {Taruya}\ \emph {et~al.}(2013)\citenamefont {Taruya},
  \citenamefont {Nishimichi},\ and\ \citenamefont
  {Bernardeau}}]{Taruya:2013my}%
  \BibitemOpen
  \bibfield  {author} {\bibinfo {author} {\bibfnamefont {A.}~\bibnamefont
  {Taruya}}, \bibinfo {author} {\bibfnamefont {T.}~\bibnamefont {Nishimichi}},
  \ and\ \bibinfo {author} {\bibfnamefont {F.}~\bibnamefont {Bernardeau}},\
  }\href {\doibase 10.1103/PhysRevD.87.083509} {\bibfield  {journal} {\bibinfo
  {journal} {Phys. Rev.}\ }\textbf {\bibinfo {volume} {D87}},\ \bibinfo {pages}
  {083509} (\bibinfo {year} {2013})},\ \Eprint {http://arxiv.org/abs/1301.3624}
  {arXiv:1301.3624 [astro-ph.CO]} \BibitemShut {NoStop}%
\bibitem [{\citenamefont {Ishikawa}\ \emph {et~al.}(2014)\citenamefont
  {Ishikawa}, \citenamefont {Totani}, \citenamefont {Nishimichi}, \citenamefont
  {Takahashi}, \citenamefont {Yoshida},\ and\ \citenamefont
  {Tonegawa}}]{Ishikawa:2013aea}%
  \BibitemOpen
  \bibfield  {author} {\bibinfo {author} {\bibfnamefont {T.}~\bibnamefont
  {Ishikawa}}, \bibinfo {author} {\bibfnamefont {T.}~\bibnamefont {Totani}},
  \bibinfo {author} {\bibfnamefont {T.}~\bibnamefont {Nishimichi}}, \bibinfo
  {author} {\bibfnamefont {R.}~\bibnamefont {Takahashi}}, \bibinfo {author}
  {\bibfnamefont {N.}~\bibnamefont {Yoshida}}, \ and\ \bibinfo {author}
  {\bibfnamefont {M.}~\bibnamefont {Tonegawa}},\ }\href {\doibase
  10.1093/mnras/stu1382} {\bibfield  {journal} {\bibinfo  {journal} {Mon. Not.
  Roy. Astron. Soc.}\ }\textbf {\bibinfo {volume} {443}},\ \bibinfo {pages}
  {3359} (\bibinfo {year} {2014})},\ \Eprint {http://arxiv.org/abs/1308.6087}
  {arXiv:1308.6087 [astro-ph.CO]} \BibitemShut {NoStop}%
\bibitem [{\citenamefont {Song}\ \emph {et~al.}(2015)\citenamefont {Song},
  \citenamefont {Taruya}, \citenamefont {Linder}, \citenamefont {Koyama},
  \citenamefont {Sabiu}, \citenamefont {Zhao}, \citenamefont {Bernardeau},
  \citenamefont {Nishimichi},\ and\ \citenamefont {Okumura}}]{Song:2015oza}%
  \BibitemOpen
  \bibfield  {author} {\bibinfo {author} {\bibfnamefont {Y.-S.}\ \bibnamefont
  {Song}}, \bibinfo {author} {\bibfnamefont {A.}~\bibnamefont {Taruya}},
  \bibinfo {author} {\bibfnamefont {E.}~\bibnamefont {Linder}}, \bibinfo
  {author} {\bibfnamefont {K.}~\bibnamefont {Koyama}}, \bibinfo {author}
  {\bibfnamefont {C.~G.}\ \bibnamefont {Sabiu}}, \bibinfo {author}
  {\bibfnamefont {G.-B.}\ \bibnamefont {Zhao}}, \bibinfo {author}
  {\bibfnamefont {F.}~\bibnamefont {Bernardeau}}, \bibinfo {author}
  {\bibfnamefont {T.}~\bibnamefont {Nishimichi}}, \ and\ \bibinfo {author}
  {\bibfnamefont {T.}~\bibnamefont {Okumura}},\ }\href {\doibase
  10.1103/PhysRevD.92.043522} {\bibfield  {journal} {\bibinfo  {journal} {Phys.
  Rev.}\ }\textbf {\bibinfo {volume} {D92}},\ \bibinfo {pages} {043522}
  (\bibinfo {year} {2015})},\ \Eprint {http://arxiv.org/abs/1507.01592}
  {arXiv:1507.01592 [astro-ph.CO]} \BibitemShut {NoStop}%
\bibitem [{\citenamefont {Beutler}\ \emph {et~al.}(2016)\citenamefont {Beutler}
  \emph {et~al.}}]{Beutler:2016arn}%
  \BibitemOpen
  \bibfield  {author} {\bibinfo {author} {\bibfnamefont {F.}~\bibnamefont
  {Beutler}} \emph {et~al.} (\bibinfo {collaboration} {BOSS}),\ }\href@noop {}
  {\bibfield  {journal} {\bibinfo  {journal} {Submitted to: Mon. Not. Roy.
  Astron. Soc.}\ } (\bibinfo {year} {2016})},\ \Eprint
  {http://arxiv.org/abs/1607.03150} {arXiv:1607.03150 [astro-ph.CO]}
  \BibitemShut {NoStop}%
\bibitem [{\citenamefont {Bose}\ and\ \citenamefont
  {Koyama}(2016)}]{Bose:2016qun}%
  \BibitemOpen
  \bibfield  {author} {\bibinfo {author} {\bibfnamefont {B.}~\bibnamefont
  {Bose}}\ and\ \bibinfo {author} {\bibfnamefont {K.}~\bibnamefont {Koyama}},\
  }\href {\doibase 10.1088/1475-7516/2016/08/032} {\bibfield  {journal}
  {\bibinfo  {journal} {JCAP}\ }\textbf {\bibinfo {volume} {1608}},\ \bibinfo
  {pages} {032} (\bibinfo {year} {2016})},\ \Eprint
  {http://arxiv.org/abs/1606.02520} {arXiv:1606.02520 [astro-ph.CO]}
  \BibitemShut {NoStop}%
\bibitem [{\citenamefont {Barreira}\ \emph {et~al.}(2016)\citenamefont
  {Barreira}, \citenamefont {Sanchez},\ and\ \citenamefont
  {Schmidt}}]{Barreira:2016mg}%
  \BibitemOpen
  \bibfield  {author} {\bibinfo {author} {\bibfnamefont {A.}~\bibnamefont
  {Barreira}}, \bibinfo {author} {\bibfnamefont {A.~G.}\ \bibnamefont
  {Sanchez}}, \ and\ \bibinfo {author} {\bibfnamefont {F.}~\bibnamefont
  {Schmidt}},\ }\href@noop {} {\  (\bibinfo {year} {2016})},\ \Eprint
  {http://arxiv.org/abs/1605.03965} {arXiv:1605.03965 [astro-ph.CO]}
  \BibitemShut {NoStop}%
\bibitem [{\citenamefont {Bose}\ \emph {et~al.}(2017)\citenamefont {Bose},
  \citenamefont {Koyama}, \citenamefont {Hellwing}, \citenamefont {Zhao},\ and\
  \citenamefont {Winther}}]{Bose:2017myh}%
  \BibitemOpen
  \bibfield  {author} {\bibinfo {author} {\bibfnamefont {B.}~\bibnamefont
  {Bose}}, \bibinfo {author} {\bibfnamefont {K.}~\bibnamefont {Koyama}},
  \bibinfo {author} {\bibfnamefont {W.~A.}\ \bibnamefont {Hellwing}}, \bibinfo
  {author} {\bibfnamefont {G.-B.}\ \bibnamefont {Zhao}}, \ and\ \bibinfo
  {author} {\bibfnamefont {H.~A.}\ \bibnamefont {Winther}},\ }\href {\doibase
  10.1103/PhysRevD.96.023519} {\bibfield  {journal} {\bibinfo  {journal} {Phys.
  Rev.}\ }\textbf {\bibinfo {volume} {D96}},\ \bibinfo {pages} {023519}
  (\bibinfo {year} {2017})},\ \Eprint {http://arxiv.org/abs/1702.02348}
  {arXiv:1702.02348 [astro-ph.CO]} \BibitemShut {NoStop}%
\bibitem [{\citenamefont {Baumann}\ \emph {et~al.}(2012)\citenamefont
  {Baumann}, \citenamefont {Nicolis}, \citenamefont {Senatore},\ and\
  \citenamefont {Zaldarriaga}}]{Baumann:2010tm}%
  \BibitemOpen
  \bibfield  {author} {\bibinfo {author} {\bibfnamefont {D.}~\bibnamefont
  {Baumann}}, \bibinfo {author} {\bibfnamefont {A.}~\bibnamefont {Nicolis}},
  \bibinfo {author} {\bibfnamefont {L.}~\bibnamefont {Senatore}}, \ and\
  \bibinfo {author} {\bibfnamefont {M.}~\bibnamefont {Zaldarriaga}},\ }\href
  {\doibase 10.1088/1475-7516/2012/07/051} {\bibfield  {journal} {\bibinfo
  {journal} {JCAP}\ }\textbf {\bibinfo {volume} {1207}},\ \bibinfo {pages}
  {051} (\bibinfo {year} {2012})},\ \Eprint {http://arxiv.org/abs/1004.2488}
  {arXiv:1004.2488 [astro-ph.CO]} \BibitemShut {NoStop}%
\bibitem [{\citenamefont {Carrasco}\ \emph {et~al.}(2012)\citenamefont
  {Carrasco}, \citenamefont {Hertzberg},\ and\ \citenamefont
  {Senatore}}]{Carrasco:2012cv}%
  \BibitemOpen
  \bibfield  {author} {\bibinfo {author} {\bibfnamefont {J.~J.~M.}\
  \bibnamefont {Carrasco}}, \bibinfo {author} {\bibfnamefont {M.~P.}\
  \bibnamefont {Hertzberg}}, \ and\ \bibinfo {author} {\bibfnamefont
  {L.}~\bibnamefont {Senatore}},\ }\href {\doibase 10.1007/JHEP09(2012)082}
  {\bibfield  {journal} {\bibinfo  {journal} {JHEP}\ }\textbf {\bibinfo
  {volume} {09}},\ \bibinfo {pages} {082} (\bibinfo {year} {2012})},\ \Eprint
  {http://arxiv.org/abs/1206.2926} {arXiv:1206.2926 [astro-ph.CO]} \BibitemShut
  {NoStop}%
\bibitem [{\citenamefont {Lewandowski}\ \emph {et~al.}(2015)\citenamefont
  {Lewandowski}, \citenamefont {Senatore}, \citenamefont {Prada}, \citenamefont
  {Zhao},\ and\ \citenamefont {Chuang}}]{Lewandowski:2015ziq}%
  \BibitemOpen
  \bibfield  {author} {\bibinfo {author} {\bibfnamefont {M.}~\bibnamefont
  {Lewandowski}}, \bibinfo {author} {\bibfnamefont {L.}~\bibnamefont
  {Senatore}}, \bibinfo {author} {\bibfnamefont {F.}~\bibnamefont {Prada}},
  \bibinfo {author} {\bibfnamefont {C.}~\bibnamefont {Zhao}}, \ and\ \bibinfo
  {author} {\bibfnamefont {C.-H.}\ \bibnamefont {Chuang}},\ }\href@noop {} {\
  (\bibinfo {year} {2015})},\ \Eprint {http://arxiv.org/abs/1512.06831}
  {arXiv:1512.06831 [astro-ph.CO]} \BibitemShut {NoStop}%
\bibitem [{\citenamefont {Perko}\ \emph {et~al.}(2016)\citenamefont {Perko},
  \citenamefont {Senatore}, \citenamefont {Jennings},\ and\ \citenamefont
  {Wechsler}}]{Perko:2016puo}%
  \BibitemOpen
  \bibfield  {author} {\bibinfo {author} {\bibfnamefont {A.}~\bibnamefont
  {Perko}}, \bibinfo {author} {\bibfnamefont {L.}~\bibnamefont {Senatore}},
  \bibinfo {author} {\bibfnamefont {E.}~\bibnamefont {Jennings}}, \ and\
  \bibinfo {author} {\bibfnamefont {R.~H.}\ \bibnamefont {Wechsler}},\
  }\href@noop {} {\  (\bibinfo {year} {2016})},\ \Eprint
  {http://arxiv.org/abs/1610.09321} {arXiv:1610.09321 [astro-ph.CO]}
  \BibitemShut {NoStop}%
\bibitem [{\citenamefont {de~la Bella}\ \emph {et~al.}(2017)\citenamefont
  {de~la Bella}, \citenamefont {Regan}, \citenamefont {Seery},\ and\
  \citenamefont {Hotchkiss}}]{delaBella:2017qjy}%
  \BibitemOpen
  \bibfield  {author} {\bibinfo {author} {\bibfnamefont {L.~F.}\ \bibnamefont
  {de~la Bella}}, \bibinfo {author} {\bibfnamefont {D.}~\bibnamefont {Regan}},
  \bibinfo {author} {\bibfnamefont {D.}~\bibnamefont {Seery}}, \ and\ \bibinfo
  {author} {\bibfnamefont {S.}~\bibnamefont {Hotchkiss}},\ }\href@noop {} {\
  (\bibinfo {year} {2017})},\ \Eprint {http://arxiv.org/abs/1704.05309}
  {arXiv:1704.05309 [astro-ph.CO]} \BibitemShut {NoStop}%
\bibitem [{\citenamefont {Skordis}\ \emph {et~al.}(2015)\citenamefont
  {Skordis}, \citenamefont {Pourtsidou},\ and\ \citenamefont
  {Copeland}}]{Skordis:2015yra}%
  \BibitemOpen
  \bibfield  {author} {\bibinfo {author} {\bibfnamefont {C.}~\bibnamefont
  {Skordis}}, \bibinfo {author} {\bibfnamefont {A.}~\bibnamefont {Pourtsidou}},
  \ and\ \bibinfo {author} {\bibfnamefont {E.~J.}\ \bibnamefont {Copeland}},\
  }\href {\doibase 10.1103/PhysRevD.91.083537} {\bibfield  {journal} {\bibinfo
  {journal} {Phys. Rev.}\ }\textbf {\bibinfo {volume} {D91}},\ \bibinfo {pages}
  {083537} (\bibinfo {year} {2015})},\ \Eprint
  {http://arxiv.org/abs/1502.07297} {arXiv:1502.07297 [astro-ph.CO]}
  \BibitemShut {NoStop}%
\bibitem [{\citenamefont {Chevallier}\ and\ \citenamefont
  {Polarski}(2001)}]{Chevallier:2000qy}%
  \BibitemOpen
  \bibfield  {author} {\bibinfo {author} {\bibfnamefont {M.}~\bibnamefont
  {Chevallier}}\ and\ \bibinfo {author} {\bibfnamefont {D.}~\bibnamefont
  {Polarski}},\ }\href {\doibase 10.1142/S0218271801000822} {\bibfield
  {journal} {\bibinfo  {journal} {Int. J. Mod. Phys.}\ }\textbf {\bibinfo
  {volume} {D10}},\ \bibinfo {pages} {213} (\bibinfo {year} {2001})},\ \Eprint
  {http://arxiv.org/abs/gr-qc/0009008} {arXiv:gr-qc/0009008 [gr-qc]}
  \BibitemShut {NoStop}%
\bibitem [{\citenamefont {Linder}(2003)}]{Linder:2002et}%
  \BibitemOpen
  \bibfield  {author} {\bibinfo {author} {\bibfnamefont {E.~V.}\ \bibnamefont
  {Linder}},\ }\href {\doibase 10.1103/PhysRevLett.90.091301} {\bibfield
  {journal} {\bibinfo  {journal} {Phys. Rev. Lett.}\ }\textbf {\bibinfo
  {volume} {90}},\ \bibinfo {pages} {091301} (\bibinfo {year} {2003})},\
  \Eprint {http://arxiv.org/abs/astro-ph/0208512} {arXiv:astro-ph/0208512
  [astro-ph]} \BibitemShut {NoStop}%
\bibitem [{\citenamefont {Jaber}\ and\ \citenamefont {de~la
  Macorra}(2016)}]{Jaber:2016ucq}%
  \BibitemOpen
  \bibfield  {author} {\bibinfo {author} {\bibfnamefont {M.}~\bibnamefont
  {Jaber}}\ and\ \bibinfo {author} {\bibfnamefont {A.}~\bibnamefont {de~la
  Macorra}},\ }\href@noop {} {\  (\bibinfo {year} {2016})},\ \Eprint
  {http://arxiv.org/abs/1604.01442} {arXiv:1604.01442 [astro-ph.CO]}
  \BibitemShut {NoStop}%
\bibitem [{\citenamefont {Koyama}\ \emph {et~al.}(2009)\citenamefont {Koyama},
  \citenamefont {Taruya},\ and\ \citenamefont {Hiramatsu}}]{Koyama:2009me}%
  \BibitemOpen
  \bibfield  {author} {\bibinfo {author} {\bibfnamefont {K.}~\bibnamefont
  {Koyama}}, \bibinfo {author} {\bibfnamefont {A.}~\bibnamefont {Taruya}}, \
  and\ \bibinfo {author} {\bibfnamefont {T.}~\bibnamefont {Hiramatsu}},\ }\href
  {\doibase 10.1103/PhysRevD.79.123512} {\bibfield  {journal} {\bibinfo
  {journal} {Phys.Rev.}\ }\textbf {\bibinfo {volume} {D79}},\ \bibinfo {pages}
  {123512} (\bibinfo {year} {2009})},\ \Eprint {http://arxiv.org/abs/0902.0618}
  {arXiv:0902.0618 [astro-ph.CO]} \BibitemShut {NoStop}%
\bibitem [{\citenamefont {Taruya}(2016)}]{Taruya:2016jdt}%
  \BibitemOpen
  \bibfield  {author} {\bibinfo {author} {\bibfnamefont {A.}~\bibnamefont
  {Taruya}},\ }\href {\doibase 10.1103/PhysRevD.94.023504} {\bibfield
  {journal} {\bibinfo  {journal} {Phys. Rev.}\ }\textbf {\bibinfo {volume}
  {D94}},\ \bibinfo {pages} {023504} (\bibinfo {year} {2016})},\ \Eprint
  {http://arxiv.org/abs/1606.02168} {arXiv:1606.02168 [astro-ph.CO]}
  \BibitemShut {NoStop}%
\bibitem [{\citenamefont {Jeong}\ and\ \citenamefont
  {Komatsu}(2006)}]{Jeong:2006xd}%
  \BibitemOpen
  \bibfield  {author} {\bibinfo {author} {\bibfnamefont {D.}~\bibnamefont
  {Jeong}}\ and\ \bibinfo {author} {\bibfnamefont {E.}~\bibnamefont
  {Komatsu}},\ }\href {\doibase 10.1086/507781} {\bibfield  {journal} {\bibinfo
   {journal} {Astrophys. J.}\ }\textbf {\bibinfo {volume} {651}},\ \bibinfo
  {pages} {619} (\bibinfo {year} {2006})},\ \Eprint
  {http://arxiv.org/abs/astro-ph/0604075} {arXiv:astro-ph/0604075 [astro-ph]}
  \BibitemShut {NoStop}%
\bibitem [{\citenamefont {Carlson}\ \emph {et~al.}(2009)\citenamefont
  {Carlson}, \citenamefont {White},\ and\ \citenamefont
  {Padmanabhan}}]{Carlson:2009it}%
  \BibitemOpen
  \bibfield  {author} {\bibinfo {author} {\bibfnamefont {J.}~\bibnamefont
  {Carlson}}, \bibinfo {author} {\bibfnamefont {M.}~\bibnamefont {White}}, \
  and\ \bibinfo {author} {\bibfnamefont {N.}~\bibnamefont {Padmanabhan}},\
  }\href {\doibase 10.1103/PhysRevD.80.043531} {\bibfield  {journal} {\bibinfo
  {journal} {Phys. Rev.}\ }\textbf {\bibinfo {volume} {D80}},\ \bibinfo {pages}
  {043531} (\bibinfo {year} {2009})},\ \Eprint {http://arxiv.org/abs/0905.0479}
  {arXiv:0905.0479 [astro-ph.CO]} \BibitemShut {NoStop}%
\bibitem [{\citenamefont {Scoccimarro}(2004)}]{Scoccimarro:2004tg}%
  \BibitemOpen
  \bibfield  {author} {\bibinfo {author} {\bibfnamefont {R.}~\bibnamefont
  {Scoccimarro}},\ }\href {\doibase 10.1103/PhysRevD.70.083007} {\bibfield
  {journal} {\bibinfo  {journal} {Phys.Rev.}\ }\textbf {\bibinfo {volume}
  {D70}},\ \bibinfo {pages} {083007} (\bibinfo {year} {2004})},\ \Eprint
  {http://arxiv.org/abs/astro-ph/0407214} {arXiv:astro-ph/0407214 [astro-ph]}
  \BibitemShut {NoStop}%
\bibitem [{\citenamefont {Percival}\ and\ \citenamefont
  {White}(2009)}]{Percival:2008sh}%
  \BibitemOpen
  \bibfield  {author} {\bibinfo {author} {\bibfnamefont {W.~J.}\ \bibnamefont
  {Percival}}\ and\ \bibinfo {author} {\bibfnamefont {M.}~\bibnamefont
  {White}},\ }\href {\doibase 10.1111/j.1365-2966.2008.14211.x} {\bibfield
  {journal} {\bibinfo  {journal} {Mon. Not. Roy. Astron. Soc.}\ }\textbf
  {\bibinfo {volume} {393}},\ \bibinfo {pages} {297} (\bibinfo {year}
  {2009})},\ \Eprint {http://arxiv.org/abs/0808.0003} {arXiv:0808.0003
  [astro-ph]} \BibitemShut {NoStop}%
\bibitem [{\citenamefont {Cole}\ \emph {et~al.}(1995)\citenamefont {Cole},
  \citenamefont {Fisher},\ and\ \citenamefont {Weinberg}}]{Cole:1994wf}%
  \BibitemOpen
  \bibfield  {author} {\bibinfo {author} {\bibfnamefont {S.}~\bibnamefont
  {Cole}}, \bibinfo {author} {\bibfnamefont {K.~B.}\ \bibnamefont {Fisher}}, \
  and\ \bibinfo {author} {\bibfnamefont {D.~H.}\ \bibnamefont {Weinberg}},\
  }\href {\doibase 10.1093/mnras/275.2.515} {\bibfield  {journal} {\bibinfo
  {journal} {Mon. Not. Roy. Astron. Soc.}\ }\textbf {\bibinfo {volume} {275}},\
  \bibinfo {pages} {515} (\bibinfo {year} {1995})},\ \Eprint
  {http://arxiv.org/abs/astro-ph/9412062} {arXiv:astro-ph/9412062 [astro-ph]}
  \BibitemShut {NoStop}%
\bibitem [{\citenamefont {Peacock}\ and\ \citenamefont
  {Dodds}(1994)}]{Peacock:1993xg}%
  \BibitemOpen
  \bibfield  {author} {\bibinfo {author} {\bibfnamefont {J.~A.}\ \bibnamefont
  {Peacock}}\ and\ \bibinfo {author} {\bibfnamefont {S.~J.}\ \bibnamefont
  {Dodds}},\ }\href {\doibase 10.1093/mnras/267.4.1020} {\bibfield  {journal}
  {\bibinfo  {journal} {Mon. Not. Roy. Astron. Soc.}\ }\textbf {\bibinfo
  {volume} {267}},\ \bibinfo {pages} {1020} (\bibinfo {year} {1994})},\ \Eprint
  {http://arxiv.org/abs/astro-ph/9311057} {arXiv:astro-ph/9311057 [astro-ph]}
  \BibitemShut {NoStop}%
\bibitem [{\citenamefont {Park}\ \emph {et~al.}(1994)\citenamefont {Park},
  \citenamefont {Vogeley}, \citenamefont {Geller},\ and\ \citenamefont
  {Huchra}}]{Park:1994fa}%
  \BibitemOpen
  \bibfield  {author} {\bibinfo {author} {\bibfnamefont {C.}~\bibnamefont
  {Park}}, \bibinfo {author} {\bibfnamefont {M.~S.}\ \bibnamefont {Vogeley}},
  \bibinfo {author} {\bibfnamefont {M.~J.}\ \bibnamefont {Geller}}, \ and\
  \bibinfo {author} {\bibfnamefont {J.~P.}\ \bibnamefont {Huchra}},\ }\href
  {\doibase 10.1086/174508} {\bibfield  {journal} {\bibinfo  {journal}
  {Astrophys. J.}\ }\textbf {\bibinfo {volume} {431}},\ \bibinfo {pages} {569}
  (\bibinfo {year} {1994})}\BibitemShut {NoStop}%
\bibitem [{\citenamefont {Ballinger}\ \emph {et~al.}(1996)\citenamefont
  {Ballinger}, \citenamefont {Peacock},\ and\ \citenamefont
  {Heavens}}]{Ballinger:1996cd}%
  \BibitemOpen
  \bibfield  {author} {\bibinfo {author} {\bibfnamefont {W.~E.}\ \bibnamefont
  {Ballinger}}, \bibinfo {author} {\bibfnamefont {J.~A.}\ \bibnamefont
  {Peacock}}, \ and\ \bibinfo {author} {\bibfnamefont {A.~F.}\ \bibnamefont
  {Heavens}},\ }\href {\doibase 10.1093/mnras/282.3.877} {\bibfield  {journal}
  {\bibinfo  {journal} {Mon. Not. Roy. Astron. Soc.}\ }\textbf {\bibinfo
  {volume} {282}},\ \bibinfo {pages} {877} (\bibinfo {year} {1996})},\ \Eprint
  {http://arxiv.org/abs/astro-ph/9605017} {arXiv:astro-ph/9605017 [astro-ph]}
  \BibitemShut {NoStop}%
\bibitem [{\citenamefont {Magira}\ \emph {et~al.}(2000)\citenamefont {Magira},
  \citenamefont {Jing},\ and\ \citenamefont {Suto}}]{Magira:1999bn}%
  \BibitemOpen
  \bibfield  {author} {\bibinfo {author} {\bibfnamefont {H.}~\bibnamefont
  {Magira}}, \bibinfo {author} {\bibfnamefont {Y.~P.}\ \bibnamefont {Jing}}, \
  and\ \bibinfo {author} {\bibfnamefont {Y.}~\bibnamefont {Suto}},\ }\href
  {\doibase 10.1086/308170} {\bibfield  {journal} {\bibinfo  {journal}
  {Astrophys. J.}\ }\textbf {\bibinfo {volume} {528}},\ \bibinfo {pages} {30}
  (\bibinfo {year} {2000})},\ \Eprint {http://arxiv.org/abs/astro-ph/9907438}
  {arXiv:astro-ph/9907438 [astro-ph]} \BibitemShut {NoStop}%
\bibitem [{\citenamefont {Peacock}(1992)}]{Peacock1992}%
  \BibitemOpen
  \bibfield  {author} {\bibinfo {author} {\bibfnamefont {J.}~\bibnamefont
  {Peacock}},\ }\href@noop {} {\bibfield  {journal} {\bibinfo  {journal} {Mon.
  Not. Roy. Astron. Soc.}\ }\textbf {\bibinfo {volume} {258}},\ \bibinfo
  {pages} {581} (\bibinfo {year} {1992})}\BibitemShut {NoStop}%
\bibitem [{\citenamefont {Springel}(2005)}]{Springel:2005mi}%
  \BibitemOpen
  \bibfield  {author} {\bibinfo {author} {\bibfnamefont {V.}~\bibnamefont
  {Springel}},\ }\href {\doibase 10.1111/j.1365-2966.2005.09655.x} {\bibfield
  {journal} {\bibinfo  {journal} {Mon. Not. Roy. Astron. Soc.}\ }\textbf
  {\bibinfo {volume} {364}},\ \bibinfo {pages} {1105} (\bibinfo {year}
  {2005})},\ \Eprint {http://arxiv.org/abs/astro-ph/0505010}
  {arXiv:astro-ph/0505010 [astro-ph]} \BibitemShut {NoStop}%
\bibitem [{\citenamefont {Zhao}(2014)}]{Zhao:2013dza}%
  \BibitemOpen
  \bibfield  {author} {\bibinfo {author} {\bibfnamefont {G.-B.}\ \bibnamefont
  {Zhao}},\ }\href {\doibase 10.1088/0067-0049/211/2/23} {\bibfield  {journal}
  {\bibinfo  {journal} {Astrophys. J. Suppl.}\ }\textbf {\bibinfo {volume}
  {211}},\ \bibinfo {pages} {23} (\bibinfo {year} {2014})},\ \Eprint
  {http://arxiv.org/abs/1312.1291} {arXiv:1312.1291 [astro-ph.CO]} \BibitemShut
  {NoStop}%
\bibitem [{\citenamefont {Aghamousa}\ \emph {et~al.}(2016)\citenamefont
  {Aghamousa} \emph {et~al.}}]{Aghamousa:2016zmz}%
  \BibitemOpen
  \bibfield  {author} {\bibinfo {author} {\bibfnamefont {A.}~\bibnamefont
  {Aghamousa}} \emph {et~al.} (\bibinfo {collaboration} {DESI}),\ }\href@noop
  {} {\  (\bibinfo {year} {2016})},\ \Eprint {http://arxiv.org/abs/1611.00036}
  {arXiv:1611.00036 [astro-ph.IM]} \BibitemShut {NoStop}%
\bibitem [{\citenamefont {Laureijs}\ \emph {et~al.}(2011)\citenamefont
  {Laureijs} \emph {et~al.}}]{Laureijs:2011gra}%
  \BibitemOpen
  \bibfield  {author} {\bibinfo {author} {\bibfnamefont {R.}~\bibnamefont
  {Laureijs}} \emph {et~al.} (\bibinfo {collaboration} {EUCLID}),\ }\href@noop
  {} {\  (\bibinfo {year} {2011})},\ \Eprint {http://arxiv.org/abs/1110.3193}
  {arXiv:1110.3193 [astro-ph.CO]} \BibitemShut {NoStop}%
\bibitem [{\citenamefont {Pantazis}\ \emph {et~al.}(2016)\citenamefont
  {Pantazis}, \citenamefont {Nesseris},\ and\ \citenamefont
  {Perivolaropoulos}}]{Pantazis:2016nky}%
  \BibitemOpen
  \bibfield  {author} {\bibinfo {author} {\bibfnamefont {G.}~\bibnamefont
  {Pantazis}}, \bibinfo {author} {\bibfnamefont {S.}~\bibnamefont {Nesseris}},
  \ and\ \bibinfo {author} {\bibfnamefont {L.}~\bibnamefont
  {Perivolaropoulos}},\ }\href {\doibase 10.1103/PhysRevD.93.103503} {\bibfield
   {journal} {\bibinfo  {journal} {Phys. Rev.}\ }\textbf {\bibinfo {volume}
  {D93}},\ \bibinfo {pages} {103503} (\bibinfo {year} {2016})},\ \Eprint
  {http://arxiv.org/abs/1603.02164} {arXiv:1603.02164 [astro-ph.CO]}
  \BibitemShut {NoStop}%
\bibitem [{\citenamefont {Abate}\ \emph {et~al.}(2012)\citenamefont {Abate}
  \emph {et~al.}}]{Abate:2012za}%
  \BibitemOpen
  \bibfield  {author} {\bibinfo {author} {\bibfnamefont {A.}~\bibnamefont
  {Abate}} \emph {et~al.} (\bibinfo {collaboration} {LSST Dark Energy
  Science}),\ }\href@noop {} {\  (\bibinfo {year} {2012})},\ \Eprint
  {http://arxiv.org/abs/1211.0310} {arXiv:1211.0310 [astro-ph.CO]} \BibitemShut
  {NoStop}%
\bibitem [{\citenamefont {Alam}\ \emph {et~al.}(2015)\citenamefont {Alam} \emph
  {et~al.}}]{Alam:2015mbd}%
  \BibitemOpen
  \bibfield  {author} {\bibinfo {author} {\bibfnamefont {S.}~\bibnamefont
  {Alam}} \emph {et~al.} (\bibinfo {collaboration} {SDSS-III}),\ }\href
  {\doibase 10.1088/0067-0049/219/1/12} {\bibfield  {journal} {\bibinfo
  {journal} {Astrophys. J. Suppl.}\ }\textbf {\bibinfo {volume} {219}},\
  \bibinfo {pages} {12} (\bibinfo {year} {2015})},\ \Eprint
  {http://arxiv.org/abs/1501.00963} {arXiv:1501.00963 [astro-ph.IM]}
  \BibitemShut {NoStop}%
\bibitem [{\citenamefont {Caldwell}\ \emph {et~al.}(2003)\citenamefont
  {Caldwell}, \citenamefont {Kamionkowski},\ and\ \citenamefont
  {Weinberg}}]{Caldwell:2003vq}%
  \BibitemOpen
  \bibfield  {author} {\bibinfo {author} {\bibfnamefont {R.~R.}\ \bibnamefont
  {Caldwell}}, \bibinfo {author} {\bibfnamefont {M.}~\bibnamefont
  {Kamionkowski}}, \ and\ \bibinfo {author} {\bibfnamefont {N.~N.}\
  \bibnamefont {Weinberg}},\ }\href {\doibase 10.1103/PhysRevLett.91.071301}
  {\bibfield  {journal} {\bibinfo  {journal} {Phys. Rev. Lett.}\ }\textbf
  {\bibinfo {volume} {91}},\ \bibinfo {pages} {071301} (\bibinfo {year}
  {2003})},\ \Eprint {http://arxiv.org/abs/astro-ph/0302506}
  {arXiv:astro-ph/0302506 [astro-ph]} \BibitemShut {NoStop}%
\bibitem [{\citenamefont {Howlett}\ \emph {et~al.}(2015)\citenamefont
  {Howlett}, \citenamefont {Manera},\ and\ \citenamefont
  {Percival}}]{Howlett:2015hfa}%
  \BibitemOpen
  \bibfield  {author} {\bibinfo {author} {\bibfnamefont {C.}~\bibnamefont
  {Howlett}}, \bibinfo {author} {\bibfnamefont {M.}~\bibnamefont {Manera}}, \
  and\ \bibinfo {author} {\bibfnamefont {W.~J.}\ \bibnamefont {Percival}},\
  }\href {\doibase 10.1016/j.ascom.2015.07.003} {\bibfield  {journal} {\bibinfo
   {journal} {Astron. Comput.}\ }\textbf {\bibinfo {volume} {12}},\ \bibinfo
  {pages} {109} (\bibinfo {year} {2015})},\ \Eprint
  {http://arxiv.org/abs/1506.03737} {arXiv:1506.03737 [astro-ph.CO]}
  \BibitemShut {NoStop}%
\bibitem [{\citenamefont {Winther}\ \emph {et~al.}(2017)\citenamefont
  {Winther}, \citenamefont {Koyama}, \citenamefont {Manera}, \citenamefont
  {Wright},\ and\ \citenamefont {Zhao}}]{Winther:2017jof}%
  \BibitemOpen
  \bibfield  {author} {\bibinfo {author} {\bibfnamefont {H.~A.}\ \bibnamefont
  {Winther}}, \bibinfo {author} {\bibfnamefont {K.}~\bibnamefont {Koyama}},
  \bibinfo {author} {\bibfnamefont {M.}~\bibnamefont {Manera}}, \bibinfo
  {author} {\bibfnamefont {B.~S.}\ \bibnamefont {Wright}}, \ and\ \bibinfo
  {author} {\bibfnamefont {G.-B.}\ \bibnamefont {Zhao}},\ }\href@noop {} {\
  (\bibinfo {year} {2017})},\ \Eprint {http://arxiv.org/abs/1703.00879}
  {arXiv:1703.00879 [astro-ph.CO]} \BibitemShut {NoStop}%
\end{thebibliography}%
\end{document}